\documentclass[journal]{IEEEtran}

\usepackage{cite}      

\usepackage{graphicx}  

\usepackage{psfrag}    

\usepackage{subfigure} 

\usepackage{url}       

\usepackage{amssymb}
\usepackage{amsmath}   
\usepackage{array}
\newtheorem{definition}{Definition}[section]
\newtheorem{theorem}{Theorem}[section]

\newtheorem{lemma}{Lemma}[section]
\newtheorem{conjecture}{Conjecture}[section]
\newenvironment{con_empty}[1]{\noindent \\ {\em Conjecture #1}}{ \par}
\newenvironment{theo_empty}[1]{\noindent \\ {\em Theorem #1}}{ \par}
\newenvironment{prop_empty}[1]{\noindent \\ {\em Proposition #1}}{ \par}
\newtheorem{proposition}{Proposition}[section]
\newtheorem{remark}{Remark}[section]

\newcommand{\rspace}{\vspace{-1.7\baselineskip}\\}

\begin{document}
%
\title{n-Channel Entropy-Constrained Multiple-Description Lattice Vector Quantization}
%
%
\author{Jan~\O stergaard,~\IEEEmembership{Student~Member,~IEEE,}
        Jesper~Jensen, and Richard~Heusdens
\thanks{This work was presented in part at the IEEE Data Compression Conference, Snowbird, Utah, 2005. This research is supported by the Technology Foundation STW, applied science division of NWO and the technology programme of the ministry of Economics Affairs. The authors are with the Department of Information and Communication Theory, Delft University of Technology, 2628 CD Delft, The Netherlands.}}%
%
%
%
\markboth{IEEE Transactions on Information Theory}{\O stergaard \MakeLowercase{\textit{et al.}}: n-Channel Entropy Constrained Multiple Description Lattice Vector Quantization}
%



\maketitle

\begin{abstract}
In this paper we derive analytical expressions for the central and side quantizers which, under high-resolutions assumptions, minimize the expected distortion of a symmetric multiple-description lattice vector quantization (MD-LVQ) system subject to entropy constraints on the side descriptions for given packet-loss probabilities. 

We consider a special case of the general $n$-channel symmetric multiple-description problem where only a single parameter controls the redundancy tradeoffs between the central and the side distortions. 
Previous work on two-channel MD-LVQ showed that the distortions of the side quantizers can be expressed through the normalized second moment of a sphere.
We show here that this is also the case for three-channel MD-LVQ\@. Furthermore, we conjecture that this is true for the general $n$-channel MD-LVQ.

For given source, target rate and packet-loss probabilities we find the optimal number of descriptions and construct the MD-LVQ system that minimizes the expected distortion. We verify theoretical expressions by numerical simulations and show in a practical setup that significant performance improvements can be achieved over state-of-the-art two-channel MD-LVQ by using three-channel MD-LVQ.
 \end{abstract}



%
\IEEEpeerreviewmaketitle

\section{Introduction}
\begin{figure}[ht]
\psfrag{D0}{$d_0$}
\psfrag{D1}{$d_1$}
\psfrag{Dc}{$d_c$}
\psfrag{R0}{$R_0$}
\psfrag{R1}{$R_1$}
\begin{center}
\includegraphics[width=7cm]{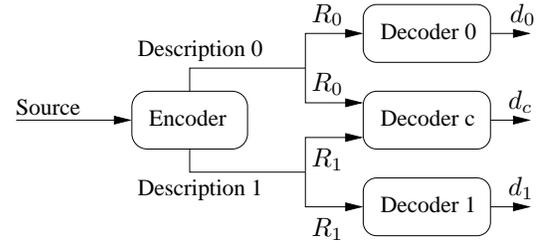}
\caption{The traditional two channel MDC scheme.}
\label{fig:twochannel}
\end{center}
\end{figure}

%
%
%
%
\PARstart{M}{ultiple} 
description coding (MDC) aims at creating separate descriptions individually capable of reproducing a source to a specified accuracy and when combined being able to refine each other. The classical scheme involves two descriptions, see Fig.~\ref{fig:twochannel}. The total rate $R$ is split between the two descriptions, i.e.\ $R=R_0+R_1$, and the distortion observed at the receiver depends on which descriptions arrive. If both descriptions are received, the distortion $(d_c)$ is lower than if only a single description is received ($d_0$ or $d_1)$.

Existing MDC schemes can roughly be divided into three categories: quantizer-based, transform-based and source-channel erasure codes based. 
Quantizer-based schemes include scalar quantization~\cite{vaishampayan:1993,vaishampayan:1994,berger-wolf:2002,tian:2004b}, trellis coded quantization~\cite{vaishampayan:1998c,jafarkhani:1999,wang:2000} and vector quantization~\cite{fleming:1999,gortz:2003,cardinal:2004,servetto:1999,diggavi:2000,vaishampayan:2001,diggavi:2002,tian:2004c,ostergaard:2005c}. Transform-based approaches include correlating transforms~\cite{orchard:1997,wang:1997,goyal:2001} and overcomplete expansions~\cite{balan:2000,chou:1999,kovacevic:2002}. Recently, schemes based on source-channel erasure codes have been introduced~\cite{pradhan:2004,pradhan:2001,puri:2002,pradhan:2001b}. 
For further details on many existing MDC techniques we refer to the survey article by Goyal~\cite{goyal:2001a}.
The present work is based on lattice vector quantization and belongs therefore to the first of the categories mentioned above.

The achievable rate-distortion (R-D) region for the two-channel problem with respect to the Gaussian source and mean-square error fidelity criterion has been known for at least two decades~\cite{ozarow:1980,elgamal:1982}. The procedures leading to the achievable region were however non-constructive, and the puzzle of designing a system capable of achieving the performance promised by theory remained unsolved. 
In 1993 Vaishampayan designed a practical MDC scheme for the scalar case~\cite{vaishampayan:1993}. The idea was to quantize the source by a central quantizer and then apply an index-assignment algorithm that uniquely mapped all reconstruction points of the central quantizer to reconstruction points in two side quantizers, thereby obtaining two coarser descriptions of the source. 
If both descriptions were received, the inverse map was applied and the performance of the central quantizer was achie\-ved, whereas if only one of the descriptions was received the source was reproduced at the resolution of one of the side quantizers. The scheme developed in~\cite{vaishampayan:1993} was, however, $8.29$ dB from the lower bound on the MDC distortion product for Gaussian sources~\cite{vaishampayan:1998,tian:2004}. Later, Vaishampayan et~al.\ described an entropy-constrained multiple-description scalar quantization system~\cite{vaishampayan:1994} that, under high-resolution assumptions, is $2.67$ dB from the lower bound~\cite{vaishampayan:1998,tian:2004}.

Recently, practical schemes for two descriptions have been introduced~\cite{servetto:1999,diggavi:2000,vaishampayan:2001,diggavi:2002}, that in the limit of infinite-dimen\-sional source vectors approach the lower bound.
Similar to \cite{vaishampayan:1993,vaishampayan:1994}, these schemes exploit the idea of having only one central quantizer followed by an index-assignment algorithm that maps each central quantizer reconstruction point to pairs of side quantizer reconstruction points. The quantizers used in~\cite{servetto:1999,diggavi:2000,vaishampayan:2001,diggavi:2002} are all lattice vector quantizers. It is common to distinguish between symmetric and asymmetric MDC\@. In the symmetric case the entropies of the side descriptions are equal and the distortions of the side descriptions are also equal whereas in the asymmetric case entropies and  distortions are allowed to be unequal. 
Multiple-description lattice vector quantization (MD-LVQ) for the symmetric case was first considered in~\cite{servetto:1999,vaishampayan:2001} where for 
given target entropies of the side descriptions $(R_0,R_1=R_0)$ as well as maximum allowable distortions $(d_0,d_1=d_0)$ of the side descriptions the central distortion $d_c$ is minimized.
It is shown that exploiting the structure in lattices makes it possible to consider only a limited region of the lattices, which makes the solution computationally feasible without sacrificing optimality. A key observation in~\cite{servetto:1999,vaishampayan:2001} is that the side distortions depend on the scaling of the lattices but are independent of the specific types of lattices. In fact the side distortions can be expressed through the normalized second moment of a sphere.

Asymmetric MD-LVQ is presented in~\cite{diggavi:2000,diggavi:2002} where the central distortion $d_c$ is minimized for given target entropies $(R_0,R_1)$ and maximum allowable side distortions $(d_0,d_1)$. A property of all the schemes presented in~\cite{servetto:1999,vaishampayan:2001,diggavi:2000,diggavi:2002} is that a simple scaling of the lattices allows adaptation to changes in target entropies without the need of any iterative training procedures.
In \cite{goyal:2002,kelner:2000} it is observed that the scheme developed in~\cite{vaishampayan:2001} is not able to continuously trade off central distortion versus side distortions. However, using non-lattices obtained by slightly modifying the lattices in~\cite{vaishampayan:2001} in an iterative fashion that alternates between optimizing the encoder while keeping the decoder fixed and optimizing the decoder while keeping the encoder fixed, it is possible to obtain a continuous range of redundancies. The problem of achieving a continuous range of redundancies is treated in more detail in~\cite{tian:2004c}.

The schemes mentioned above all consider two descriptions and the extension to more than two descriptions is not straightforward. 
State-of-the-art schemes for more than two descriptions are based on source-channel erasure codes~\cite{pradhan:2004,pradhan:2001,puri:2002,pradhan:2001b} which are fundamentally different from the quantizer-based approaches considered above. Schemes based on source-channel erasure codes rely upon the assumption that at least $\kappa$ out of $K$ descriptions are received, for some pre-specified $\kappa$. If less than $\kappa$ descriptions are received, the quality of the reconstructed source is poor and if $\kappa$ or more descriptions are received a good quality can be achieved. 
Among the few quantizer-based approaches which consider more than two descriptions are~\cite{berger-wolf:2002,tian:2004b,fleming:1999,gortz:2003,cardinal:2004,ostergaard:2005c}.

In this paper\footnote{A conference version of this work appeared in~\cite{ostergaard:2005c}.} we consider a special case of the general $n$-channel symmetric multiple-description problem where only a single\footnote{We show in~\cite{ostergaard:2006b} that additional control parameters can be included in the MD-LVQ scheme presented in this paper by exploiting recent results on distributed source coding~\cite{pradhan:2004}.} parameter controls the redundancy tradeoffs between the central and the side distortions. With a single controling parameter it is possible to describe the entire symmetric R-D region for two descriptions as shown in~\cite{servetto:1999,vaishampayan:2001} but it is not enough to describe the symmetric achievable $n$-channel R-D region. As such the proposed scheme offer a partial solution to the problem of designing balanced MD-LVQ systems.

We derive analytical expressions for the central and side quantizers which, under high-resolutions assumptions, minimize the \emph{expected distortion} at the receiving side subject to entropy constraints on the side descriptions for given packet-loss probabilities. 
The central and side quantizers we use are lattice vector quantizers as presented in \cite{vaishampayan:2001,diggavi:2002}. The central distortion, in our scheme, depends upon the lattice in question whereas the side distortions only depend on the scaling of the lattices but are independent of the specific types of lattices. In the case of three descriptions we show that the side distortions can be expressed through the normalized second moment of a sphere as was the case for the two descriptions system presented in~\cite{servetto:1999,vaishampayan:2001}. Furthermore, we conjecture that this is true in the general case of an arbitrary number of descriptions.

While state-of-the-art quantizer-based MDC schemes~\cite{vaishampayan:2001,diggavi:2002} mainly deal with only two descriptions, we construct balanced quantizers for an \emph{arbitrary} number of descriptions. In the presented approach the expected distortion observed at the receiving side depends only upon the number of received descriptions, hence the descriptions are mutually refinable and reception of any $\kappa$ out of $K$ descriptions yields equivalent expected distortion. This is different from successive refinement schemes~\cite{equitz:1991} where the individual descriptions often must be received in a prescribed order to be able to refine each other, i.e.\ description number $l$ will not do any good unless descriptions $0,\dots,l-1$ have already been received. We construct a scheme which for given packet-loss probabilities and a maximum bit budget (target entropy) determines the optimal number of descriptions and specifies the quantizers that minimize the expected distortion. 

This paper is structured as follows. In Section~\ref{sec:prelim} we briefly review specific lattice properties and introduce the concept of an index-assignment algorithm. The actual design of the index-assignment algorithm is deferred to Section~\ref{sec:label}. Reconstruction of the source and optimal construction of the labeling function is also presented in Section~\ref{sec:label}. In Section~\ref{sec:highrate} we present a high-resolution analysis of the expected distortion. We describe how to construct the quantizers in Section~\ref{sec:construction} and numerical evaluation follows in Section~\ref{sec:num}. Appendices contain proofs of Theorems.

\section{Preliminaries}\label{sec:prelim}
In this work we use lattices as vector quantizers. For a general treatment of quantizers based on lattices, see~\cite{gray:1990,conway:1999,gibson:1988}. 
This section briefly review lattice properties, introduces the concept of index assignments and describe important results regarding rate and distortion performance of MD-LVQ systems.

\subsection{Lattice Properties}
A real $L$-dimensional lattice $\Lambda$ is a discrete set of points in the $L$-dimensional Euclidean space $\mathbb{R}^L$. It forms an additive group under ordinary vector addition and can be specified through $L$ independent basis vectors~\cite{ericson:2001}. The lattice then consists of all possible integral linear combinations of the basis vectors, or, more formally
\begin{equation}
\Lambda = \left\{ \lambda \in \mathbb{R}^L : \lambda = \sum_i l_i b_i,\, l_i \in \mathbb{Z}\right\},
\end{equation}
where $b_i$ are the basis vectors also known as generator vectors of the lattice.

When $\Lambda$ is used as a vector quantizer, a point (vector) $x\in \mathbb{R}^L$ is mapped to the closest lattice point $\lambda \in \Lambda$.
The lattice points are then the codewords (reproduction points) of the quantizer.
This quantization process partitions the space $\mathbb{R}^L$ into cells called Voronoi cells, Voronoi regions or nearest-neighbor decision regions. 
The Voronoi cells of a lattice are congruent polytopes\footnote{A polytope is a finite convex region enclosed by a finite number of hyperplanes~\cite{coxeter:1973}.}, hence they are similar in size and shape and may be seen as translated versions of a fundamental region, e.g.\ the Voronoi cell around origo. A Voronoi cell, $V(\lambda)$, where $\lambda\in \Lambda$, is given by
\begin{equation}
V(\lambda) \triangleq \{ x\in \mathbb{R}^L : \| x - \lambda\|^2 \leq \|x-\lambda'\|^2,\, \forall\, \lambda' \in \Lambda \},
\end{equation}
and we write $Q(x)=\lambda$ if $x\in V(\lambda)$. Throughout this work we will be considering the $l_2$-norm (normalized per dimension) given by $\|x\|^2=\langle x, x \rangle$, where the inner product is defined as
\begin{equation}
\langle x, y \rangle \triangleq \frac{1}{L}\sum_{i=0}^{L-1}x_iy_i.
\end{equation}

A lattice is completely specified by its fundamental region, and often expressed through the volume $\nu$ of the fundamental region as well as its dimensionless normalized second moment of inertia $G(\Lambda)$~\cite{conway:1999}, which is given by
\begin{equation}\label{eq:G}
G(\Lambda) \triangleq\frac{1}{\nu^{1+2/L}}\int_{V(0)}\|x\|^2dx,
\end{equation}
where $V(0)$ is the Voronoi cell around origo. Applying any scaling or orthogonal transform, e.g.\ rotation or reflection on $\Lambda$ will not change $G(\Lambda)$, which makes it a good figure of merit when comparing different lattices (quantizers). In other words, $G(\Lambda)$ depends only upon the shape of the fundamental region, and in general, the more sphere-like shape, the lower normalized second moment. 

In this paper we consider one central quantizer and $K$ side quantizers. The central quantizer is based on a central lattice $\Lambda_c\subset \mathbb{R}^L$ with fundamental regions of volume $\nu=\det(\Lambda_c)$. The side quantizers are based on a geometrical similar\footnote{A lattice $\Lambda_s$ is said to be geometrical similar to $\Lambda_c$ if $\Lambda_s$ can be obtained from $\Lambda_c$ by applying a change of scale, a rotation and possible a reflection\cite{conway:1999}.} sublattice $\Lambda_s\subseteq \Lambda_c$ of index $N=[\Lambda_c:\Lambda_s]$ and fundamental regions of volume $\nu_s=\nu N$. The trivial case $K=1$ leads to a single-description system, where we would simply use one central quantizer and no side quantizers. 

We will consider the balanced situation, where the entropy $R_s$ is the same for each description. Furthermore, we consider the case where the contribution $d_i, i=0,\dots,K-1$ of each description to the total distortion is the same. Our design makes sure\footnote{We prove this symmetry property for the asymptotical case of $N\rightarrow\infty$ and $\nu_s\rightarrow 0$. For finite $N$ we do not guarantee the existence of an exact symmetric solution. However, by use of time-sharing, it is always possible to achieve symmetry.}  that the distortion observed at the receiving side, depends only on the number of descriptions received, hence reception of any $\kappa$ out of $K$ descriptions yields equivalent expected distortion.

\subsection{Index Assignments}\label{sec:index}
In the MDC scheme considered in this paper, a source vector $x$ is quantized to the nearest reconstruction point $\lambda_c$ in the central lattice $\Lambda_c$. Hereafter follows index assignments (mappings), which uniquely maps all $\lambda_c$'s to vectors in each of the side quantizers. This mapping is done through a labeling function $\alpha$, and we denote the individual component functions of $\alpha$ by $\alpha_i$, where $i=0,\dots, K-1$. In other words, the injective map $\alpha$ that maps $\Lambda_c$ into $\Lambda_s\times\dots \times \Lambda_s$, is given by
\begin{align}
\alpha(\lambda_c)&=(\alpha_0(\lambda_c),\alpha_1(\lambda_c),\dots,\alpha_{K-1}(\lambda_c)) \\
&= (\lambda_0,\lambda_1,\dots,\lambda_{K-1}),
\end{align}
where $\alpha_i(\lambda_c)=\lambda_i \in \Lambda_s$ and $i=0,\dots, K-1$. Each $K$-tuple $(\lambda_0,\dots,\lambda_{K-1})$ is used only once when labeling points in $\Lambda_c$ in order to make sure that $\lambda_c$ can be recovered unambiguously when all $K$ descriptions are received. At this point we also define the inverse component map, $\alpha_i^{-1}$, which gives a set of central lattice points a specific sublattice point is mapped to. This is given by
\begin{equation}
\alpha_i^{-1}(\lambda_i) = \{\lambda_c \in \Lambda_c : \alpha_i(\lambda_c)=\lambda_i\}\quad \text{for all}\quad \lambda_i \in \Lambda_s,
\end{equation}
where $|\alpha_i^{-1}(\lambda_i)| \approx N$, since there are $N$ times as many central lattice points as sublattice points within a bounded region of $\mathbb{R}^L$.

Since lattices are infinite arrays of points, we construct a shift invariant labeling function, so we only need to label a finite number of points as is done in~\cite{vaishampayan:2001,diggavi:2002}. Following the approach in~\cite{diggavi:2002} we construct a product lattice $\Lambda_\pi$ which has $N^2$ central lattice points and $N$ sublattice points in each of its Voronoi cells. The Voronoi cells $V_\pi$ of the product lattice $\Lambda_\pi$ are all similar so by concentrating on labeling only central lattice points within one Voronoi cell of the product lattice, the rest of the central lattice points may be labeled simply by translating this Voronoi cell throughout $\mathbb{R}^L$. Other choices of product lattices are possible, but this choice has a particular simple construction.  
With this choice of product lattice, we only label central lattice points within $V_\pi(0)$, which is the Voronoi cell of $\Lambda_\pi$ around origo. With this we get 
\begin{equation}\label{eq:shiftinv}
\alpha(\lambda_c + \lambda_\pi) = \alpha(\lambda_c) + \lambda_\pi,
\end{equation}
for all $\lambda_\pi \in \Lambda_\pi$ and all $\lambda_c \in \Lambda_c$. 

\subsection{Rate and Distortion Performance of MD-LVQ Systems}

\subsubsection{Central Distortion}\label{sec:centraldist}
We consider a source that generates independent identically distributed random variables with probability density function (pdf) $f$. Let $X \in \mathbb{R}^L$ be a random vector made by blocking the source into vectors of length $L$, and let $x\in \mathbb{R}^L$ denote a realization of $X$. The $L$-fold pdf of $X$ is denoted $f_{X}$ and given by
\begin{equation}
f_X(x) = \prod_{j=0}^{L-1}f(x_j).
\end{equation}
The expected central distortion $d_c$ is defined as
\begin{equation}\label{eq:d0}
d_c \triangleq \sum_{\lambda_c \in \Lambda_c}\int_{V_c(\lambda_c)}\|x-\lambda_c\|^2f_X(x)dx,
\end{equation}
where $V_c(\lambda_c)$ is the Voronoi cell of a single reconstruction point $\lambda_c \in \Lambda_c$. 
Using standard high-resolution assumptions for lattice quantizers~\cite{gersho:1991,gray:1990,gibson:1988}, the expected central distortion can be expressed in terms of the dimensionless normalized second moment of inertia, $G(\Lambda_c)$, that is
\begin{equation}\label{eq:d0G}
d_c \approx G(\Lambda_c)\nu^{2/L},
\end{equation}
where $G(\Lambda_c)$ is given by~(\ref{eq:G}).

\subsubsection{Side Distortions}
The side distortion for the $i$th description is given by
\begin{equation}
d_i = \sum_{\lambda_c \in \Lambda_c}\int_{V_c(\lambda_c)} \|x - \alpha_i(\lambda_c)\|^2f_X(x)dx,\quad  i=0,\dots, K-1,
\end{equation}
which can be approximated as~\cite{vaishampayan:2001}
\begin{equation}\label{eq:di}
d_i \approx d_c + \sum_{\lambda_c \in \Lambda_c} \|\lambda_c - \alpha_i(\lambda_c)\|^2P(\lambda_c),
\end{equation}
where $P(\lambda_c)$ is the probability that $X$ will be mapped to $\lambda_c$, i.e.\ $P(Q(X)=\lambda_c)=\int_{V_c(\lambda_c)}f_X(x)\,dx$.
We notice that independent of which labeling function we use, the distortion introduced by the central quantizer is orthogonal (under high-resolution assumptions) to the distortion introduced by the side quantizers.
Exploiting the shift-invariance property of the labeling function (\ref{eq:shiftinv}) makes it possible to simplify (\ref{eq:di}) as
\begin{equation}\label{eq:di1}
\begin{split}
d_i&\approx d_c + \sum_{\lambda_\pi \in \Lambda_\pi}\frac{P(\lambda_\pi)}{N^2}
\sum_{\lambda_c \in V_\pi(0)} \|\lambda_c - \alpha_i(\lambda_c)\|^2\\
&= d_c + \frac{1}{N^2}\sum_{\lambda_c \in V_\pi(0)} \|\lambda_c - \alpha_i(\lambda_c)\|^2,\quad i=0,\dots, K-1,
\end{split}
\end{equation}
where we assume the region $V_\pi(0)$ is sufficiently small so $P(\lambda_c) \approx P(\lambda_\pi)/N^2$, for $\lambda_c\in V_\pi(\lambda_\pi)$. Notice that we assume $P(\lambda_\pi)$ to be constant only within each region $V_\pi(\lambda_\pi)$, hence it may take on different values for each $\lambda_\pi \in \Lambda_\pi$.

\subsubsection{Rate}\label{sec:rate}
\begin{definition}
$R_c = H(Q(X))/L$ denotes the minimum entropy needed for a single-description system to achieve an expected distortion of $d_c$, the central distortion of the multiple-description system as given by (\ref{eq:d0G}).
\end{definition}

The single-description rate $R_c$ is given by
\begin{equation}
R_c = -\frac{1}L\sum_{\lambda_c\in \Lambda_c}\int_{V_c(\lambda_c)}f_X(x)dx\,\log_2\left(\int_{V_c(\lambda_c)}f_X(x)dx\right).
\end{equation}
Using that each quantizer cell has identical volume $\nu$ and assuming that $f_X(x)$ is approximately constant within Voronoi cells of the central lattice $\Lambda_c$, it can be shown that
\begin{equation}\label{eq:R0}
R_c \approx h(X) - \frac{1}L\log_2(\nu),
\end{equation}
where $h(X)$ is the component-wise differential entropy of a source vector.

\begin{definition}
$R_s$ denotes the entropy of the individual descriptions in a balanced multiple-description system. The entropy of the $i$th description is given by $R_s=H(\alpha_i(Q(X)))/L$, where $i=0,\dots,K-1$.
\end{definition}

The side descriptions are based on a coarser lattice obtained by scaling the Voronoi cells of the central lattice by a factor of $N$. Assuming the pdf of $X$ is roughly constant within a sublattice cell, the entropy of the side descriptions is given by
\begin{equation}\label{eq:Rs}
R_s\approx h(X) - \frac{1}L\log_2(N\nu).
\end{equation}
The entropy of the side descriptions is related to the entropy of the single-description system by
\begin{equation}
R_s = R_c - \frac{1}L\log_2(N).
\end{equation}

\section{Construction of Labeling Function}
\label{sec:label}
The index assignment is done by a labeling function $\alpha$, that maps central lattice points to sublattice points. An optimal index assignment minimizes a cost functional when $0< \kappa < K$ descriptions are received. In addition, the index assignment should be invertible so the central quantizer can be used when all descriptions are received. Before defining the labeling function we have to define the cost functional to be minimized. To do so, we first describe how to approximate the source sequence when receiving only $\kappa$ descriptions and how to determine the expected distortion.
Then we define the cost functional to be minimized by the labeling function $\alpha$ and describe how to minimize it.

\subsection{Expected Distortion}
\label{sec:reconstruction}
At the receiving side, $X\in \mathbb{R}^L$ is reconstructed to a quality that is determined only by the number of received descriptions. If no descriptions are received we reconstruct using the expected value, $E[X]$, and if all $K$ descriptions are received we reconstruct using the inverse map $\alpha^{-1}$, hence obtaining the quality of the central quantizer. 

In this work we use a simple reconstruction rule which applies for arbitrary sources. When receiving $1\leq \kappa<K$ descriptions we reconstruct using the average of the $\kappa$ descriptions. We show later (Theorem~\ref{theo:sums}) that using the average of received descriptions as reconstruction rule makes it possible to split the distortion due to reception of any number of descriptions into a sum of squared norms between pairs of lattice points. Moreover, this lead to the fact that the side quantizers performances approach that of quantizers having spherical Voronoi regions.

There are in general several ways of receiving $\kappa$ out of $K$ descriptions. Let $\mathcal{L}$ denote an index set consisting of all possible $\kappa$ combinations out of $\{0,\dots, K-1\}$. Hence $|\mathcal{L}| = \binom{K}{\kappa}$. We denote an element of $\mathcal{L}$ by $l=\{l_0, \dots , l_{\kappa-1}\}\in \mathcal{L}$. Upon reception of any $\kappa$ descriptions we reconstruct to $\hat{X}$ using 
\begin{equation}
\hat{X}=\frac{1}{\kappa}\sum_{j=0}^{\kappa-1}\lambda_{l_j},
\end{equation}
where $l \in \mathcal{L}$.

Assuming packet-loss probabilities are independent and are the same for all descriptions, say $p$, we may write the expected distortion when receiving $\kappa$ out $K$ descriptions as
\begin{equation}\label{eq:expdist}
\begin{split}
&d_a^{(K,\kappa)}\approx (1-p)^{\kappa}p^{K-\kappa} \\
&\times \left(
\binom{K}{\kappa}d_c 
+ \frac{1}{N^2}
\sum_{l\in \mathcal{L}}\sum_{\lambda_c\in V_\pi(0)}
\left\|\lambda_c - \frac{1}{\kappa}\sum_{j=0}^{\kappa-1} \lambda_{l_j}\right\|^2\right),
\end{split}
\end{equation}
where $\lambda_{l_j}=\alpha_{l_j}(\lambda_c)$ and the two special cases $\kappa\in \{0,K\}$ are given by $d_a^{(K,0)}\approx p^K E[\|X\|^2]$ and $d_a^{(K,K)}\approx (1-p)^K d_c$.

\subsection{Cost Functional}
From (\ref{eq:expdist}) we see that the side distortion may be split into two terms, one describing the distortion occurring when the central quantizer is used on the source, and one that describes the distortion due to the index assignment. 
An optimal index assignment jointly minimizes the second term in (\ref{eq:expdist}) over all $1\leq \kappa\leq K-1$ possible descriptions.
The cost functional $\mathcal{J}$ to be minimized by the index assignment algorithm is then given by
\begin{equation}\label{eq:IAcost}
\mathcal{J}=\sum_{\kappa=1}^{K-1}J^{(K,\kappa)},
\end{equation}
where 
\begin{equation}\label{eq:costfunctional}
J^{(K,\kappa)}= \frac{(1-p)^{\kappa}p^{K-\kappa}}{N^2}
\sum_{l\in \mathcal{L}}\sum_{\lambda_c\in V_\pi(0)}
\left\| \lambda_c - \frac{1}{\kappa}\sum_{j=0}^{\kappa-1} \lambda_{l_j}\right\|^2.
\end{equation}
The cost functional should be minimized subject to an entropy constraint on the side descriptions. We remark here that the side entropies depend solely on $\nu$ and $N$ and as such not on the particular choice of $K$-tuples. In other words, for fixed $N$ and $\nu$ the index assignment problem is solved if~(\ref{eq:IAcost}) is minimized. The problem of choosing $\nu$ and $N$ such that the entropy constraint is satisfied is independent of the assignment problem and deferred to Section~\ref{sec:opt}.

The following theorem makes it possible to rewrite the cost functional in a way that brings more insight into which $K$-tuples to use.

\begin{theorem}\label{theo:sums}
For $1\leq \kappa \leq K$ we have
\begin{equation*}
\begin{split}
\sum_{l\in\mathcal{L}}&
\sum_{\lambda_c}
\left\| \lambda_c - \frac{1}{\kappa}\sum_{j=0}^{\kappa-1} \lambda_{l_j}\right\|^2 \\ 
&= \sum_{\lambda_c}\binom{K}{\kappa}\Bigg(\left\| \lambda_c - 
\frac{1}{K}\sum_{i=0}^{K-1}\lambda_i\right\|^2 \\
&\quad+ \left(\frac{K-\kappa}{K^2\kappa(K-1)}\right)
\sum_{i=0}^{K-2}\sum_{j=i+1}^{K-1}\| \lambda_i - \lambda_j\|^2\Bigg).
\end{split}
\end{equation*}
\end{theorem}

\begin{proof}
See Appendix~\ref{app:sums}.
\end{proof}
From Theorem~\ref{theo:sums} it is clear that~(\ref{eq:costfunctional}) can be written as
\begin{equation}\label{eq:costfunctional2}
\begin{split}
&J^{(K,\kappa)}= \frac{(1-p)^{\kappa}p^{K-\kappa}}{N^2}
\sum_{l\in\mathcal{L}}\sum_{\lambda_c\in V_\pi(0)}
\left\| \lambda_c - \frac{1}{\kappa}\sum_{j=0}^{\kappa-1} \lambda_{l_j}\right\|^2 \\ 
&=\frac{(1-p)^{\kappa}p^{K-\kappa}}{N^2}
\binom{K}{\kappa}\left(
\sum_{\lambda_c\in V_\pi(0)}\left\| \lambda_c - 
\frac{1}{K}\sum_{i=0}^{K-1}\lambda_i\right\|^2 \right.\\
&\quad +
\left.\sum_{\lambda_c\in V_\pi(0)}\left(
\frac{K-\kappa}{K^2\kappa(K-1)}\right)
\sum_{i=0}^{K-2}\sum_{j=i+1}^{K-1}\| \lambda_i - \lambda_j\|^2\right).
\end{split}
\end{equation}

The first term in~(\ref{eq:costfunctional2}) describes the distance from a central lattice point to the centroid of its associated $K$-tuple. The second term describes the sum of pairwise squared distances (SPSD) between elements of the $K$-tuples. In Section~\ref{sec:highrate} (Proposition~\ref{prop:growthriemann2}) we show that, under a high-resolution assumption, the second term in~(\ref{eq:costfunctional2}) is dominant, from which we conclude that 
in order to minimize~(\ref{eq:IAcost}) we have to choose
the $K$-tuples with the lowest SPSD\@. 
These $K$-tuples are then assigned to central lattice points in such a way, that the first term in~(\ref{eq:costfunctional2}) is minimized. 

Independent of the packet-loss probability, we always minimize the second term in~(\ref{eq:costfunctional2}) by using those $K$-tuples which have the smallest SPSD\@. This means that, at high resolution, the optimal $K$-tuples are independent of packet-loss probabilities and, consequently, the optimal assignment is independent\footnote{Given the central lattice and  the sublattice, the optimal assignment is independent of $p$. However, we show later that the optimal $N$ depends on $p$.} of the packet-loss probability.

\subsection{Minimizing Cost Functional}
\label{sec:ktuples}
In order to make sure that $\alpha$ is shift-invariant, we use unique $K$-tuples, i.e.\ $K$-tuples that are assigned to one central lattice point $\lambda_c\in \Lambda_c$ only. Notice that two $K$-tuples which are translates of each other by some $\lambda_\pi \in \Lambda_\pi$ must not both be assigned to central lattice points located within the same region $V_\pi(\lambda_\pi)$, since this causes assignment of the same $K$-tuples to multiple central lattice points.
The region $V_\pi(0)$ will be translated through-out $\mathbb{R}^L$ and centered at $\lambda_\pi \in \Lambda_\pi$, so there will be no overlap between neighboring regions, i.e.\ $V_\pi(\lambda'_\xi) \cap V_\pi(\lambda''_\xi) = \emptyset$, for $\lambda'_\xi,\lambda''_\xi \in \Lambda_\pi$ and $\lambda'_\xi \neq \lambda''_\xi$. 
One obvious way of avoiding assigning $K$-tuples to multiple central lattice points is then to exclusively use sublattice points located within $V_\pi(0)$. 
However, sublattice points located close to but outside $V_\pi(0)$, might be better  candidates than sublattice points within $V_\pi(0)$ when labeling central lattice points close to the boundary. A consistent way of constructing $K$-tuples, is to center a region $\tilde{V}$ at all sublattice points $\lambda_0 \in \Lambda_s \cap V_\pi(0)$, and construct $K$-tuples by combining sublattice points $\lambda_i\in \Lambda_s, i=1,\dots,K-1$ within $\tilde{V}(\lambda_0)$ in all possible ways and select the ones that minimize (\ref{eq:costfunctional2}). 
For a fixed $\lambda_i\in \Lambda_s$, the expression $\sum_{\lambda_j\in \Lambda_s \cap \tilde{V}(\lambda_i)}\|\lambda_i - \lambda_j\|^2$ is minimized when $\tilde{V}$ forms a sphere centered at $\lambda_i$. Our construction allows for $\tilde{V}$ to have an arbitrary shape, e.g.\ the shape of $V_\pi$ which is the shape used for the two-description system presented in~\cite{diggavi:2002}. However, if $\tilde{V}$ is not chosen to be a sphere, the SPSD is in general not minimized.

For each $\lambda_0\in \Lambda_s\cap V_\pi(0)$ it is possible to construct $\tilde{N}^{K-1}$ $K$-tuples, where $\tilde{N}$ is the number of sublattice points within the region $\tilde{V}$. 
This gives a total of $N\tilde{N}^{K-1}$ $K$-tuples when all $\lambda_0\in \Lambda_s \cap V_\pi(0)$ are used.
However, only $N^2$ central lattice points need to be labeled. When $K=2$, we let $\tilde{N}=N$, so the number of possible $K$-tuples is equal to $N^2$, which is exactly the number of central lattice points in $V_\pi(0)$. 
In general, for $K>2$, the volume $\tilde{\nu}$ of $\tilde{V}$ is smaller than the volume of $V_\pi(0)$ and as such $\tilde{N}<N$. We can approximate $\tilde{N}$ through the volumes $\nu_s$ and $\tilde{\nu}$, i.e.\ $\tilde{N}\approx \tilde{\nu}/\nu_s$. To justify this approximation let $\Lambda \subset \mathbb{R}^L$ be a real lattice and let $\nu=\det(\Lambda)$ be the volume of a fundamental region. Let $S(c,r)$ be a sphere in $\mathbb{R}^L$ of radius $r$ and center $c\in \mathbb{R}^L$. 
According to Gauss' counting principle, the number $A_{\mathbb{Z}}$ of integer lattice points in a convex body $\mathcal{C}$ in $\mathbb{R}^L$ equals the volume Vol$(\mathcal{C})$ of $\mathcal{C}$ with a small error term~\cite{mazo:1990}. In fact if $\mathcal{C}=S(c,r)$ then by use of a theorem due to Minkowski it can be shown that, for any $c\in\mathbb{R}^L$ and asymptotically as $r\rightarrow \infty$, $A_{\mathbb{Z}}(r)=\text{Vol}(S(c,r))=\omega_L r^L$, where $\omega_L$ is the volume of the $L$-dimensional unit sphere~\cite{fricker:1982}, see also~\cite{erdos:1989,bokowski:1973,vinograd:1963,gruber:1987,kratzel:1988}. 
It is also known that the number of lattice points $A_\Lambda(n)$ in the first $n$ shells of the lattice $\Lambda$ satisfies, asymptotically as $n\rightarrow \infty$, $A_\Lambda(n) = \omega_L n^{L/2}/\nu$~\cite{vaishampayan:2001}. 
Hence, based on the above we approximate the number of lattice points in $\tilde{V}$ by $\tilde{\nu}/\nu_s$, which is an approximation that becomes exact as the number of shells $n$ within $\tilde{V}$ goes to infinity\footnote{For the high-resolution analysis given in Section~\ref{sec:highrate} it is important that $\tilde{\nu}$ is kept small as the number of lattice points within $\tilde{V}$ goes to infinity. This is easily done by proper scaling of the lattices, i.e.\ making sure that $\nu_s\rightarrow 0$ as $N\rightarrow \infty$.} (which corresponds to $N\rightarrow\infty$). Our analysis is therefore only exact in the limiting case of $N\rightarrow \infty$. With this we can, in the asymptotical case of $N\rightarrow \infty$, lower bound $\tilde{\nu}$ by 
\begin{equation}\label{eq:vtilde}
\tilde{\nu}\geq \nu_s\, N^{1/(K-1)}.
\end{equation}
Hence, $\tilde{V}$ contains $\tilde{N}\geq N^{1/(K-1)}$ sublattice points so that the total number of possible $K$-tuples is $N\tilde{N}^{K-1}\geq N^2$.

\begin{figure}[ht]
\psfrag{sub}{$\in \Lambda_s$}
\psfrag{prod}{$\in \Lambda_\pi$}
\psfrag{vtilde}{$\tilde{V}$}
\psfrag{V0}{$V_\pi$}
\psfrag{Vn}{$V_\pi(0)$}
\psfrag{K=3}{$K=3$}
\psfrag{N=25}{$N=25$}
\psfrag{Nt=5}{$\tilde{N}=5$}
\begin{center}
\includegraphics{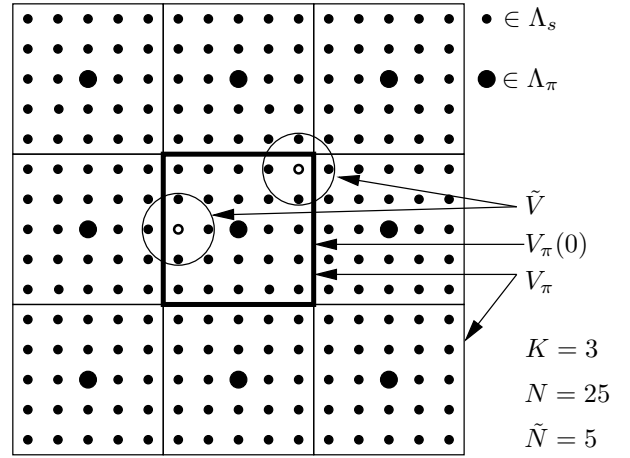}
\end{center}
\caption{The region $\tilde{V}$ is here shown centered at two different sublattice points within $V_\pi(0)$. Small dots represents sublattice points of $\Lambda_s$ and large dots represents product lattice points $\lambda_\pi \in \Lambda_\pi$. Central lattice points are not shown here. $V_\pi$ contains 25 sublattice points (shown as squares) centered at product lattice points. In this example $\tilde{V}$ contains 5 sublattice points.}
\label{fig:vtilde}
\end{figure}

In Fig.~\ref{fig:vtilde} is shown an example of $\tilde{V}$ and $V_\pi$ regions for the two-dimensional $Z^2$ lattice. In the example we used $K=3$ and $N=25$, hence there are 25 sublattice points within $V_\pi$. There are $\tilde{N}=N^{1/(K-1)}=5$ sublattice points in $\tilde{V}$ which is exactly the minimum number of points required, according to (\ref{eq:vtilde}). 

With equality in (\ref{eq:vtilde}) we obtain a region that contains the exact number of sublattice points required to construct $N$ tuples for each of the $N$ $\lambda_0$ points in $V_\pi(0)$. 
According to~(\ref{eq:costfunctional2}), a central lattice point should be assigned that $K$-tuple where a weighted average of any subset of the elements of the $K$-tuple is as close as possible to the central lattice point. 
The optimal assignment of $K$-tuples to central lattice points can be formulated and solved as a linear assignment problem~\cite{west:2001}.  

\subsubsection{Shift-Invariance by use of Cosets}
By centering $\tilde{V}$ around each $\lambda_0\in\Lambda_s\cap V_\pi(0)$, we make sure that the map $\alpha$ is shift-invariant. However, this also means that all $K$-tuples have their first coordinate (i.e.\ $\lambda_0$) inside $V_\pi(0)$. To be optimal this restriction must be removed which is easily done by considering all cosets of each $K$-tuple. The coset of a fixed $K$-tuple, say $t=(\lambda_0,\lambda_1,\dots,\lambda_{K-1})$ where $\lambda_0\in \Lambda_s\cap V_\pi(0)$ and $(\lambda_1,\dots,\lambda_{K-1}) \in (\Lambda_s\times \dots\times\Lambda_s)$, is given by $\mathrm{Coset}(t)=\{t+\lambda_\pi\}$ for all $\lambda_\pi\in\Lambda_\pi$. $K$-tuples in a coset are distinct modulo $\Lambda_\pi$ and by making sure that only one member from each coset is used, the shift-invariance property is preserved. In general it is optimal to consider only those $\lambda_\pi$ product lattice points that are close to $V_\pi(0)$, e.g.\ those points whose Voronoi cell touches $V_\pi(0)$. The number of such points is given by the kissing-number $\mathfrak{K}(\Lambda_\pi)$ of the particular lattice~\cite{conway:1999}.

\subsubsection{Dimensionless Expansion Factor $\psi_L$}
\begin{figure}[ht]
\psfrag{l0}{$\lambda_0$}
\psfrag{l1}{$\lambda_1$}
\psfrag{l2}{$\lambda_2$}
\psfrag{l1p}{$\lambda_1'$}
\psfrag{l2p}{$\lambda_2'$}
\psfrag{vtilde}{$\tilde{V}$}
\psfrag{K=3}{$K=3$}
\psfrag{N=81}{$N=81$}
\psfrag{Nt=9}{$\tilde{N}=9$}
\begin{center}
\includegraphics{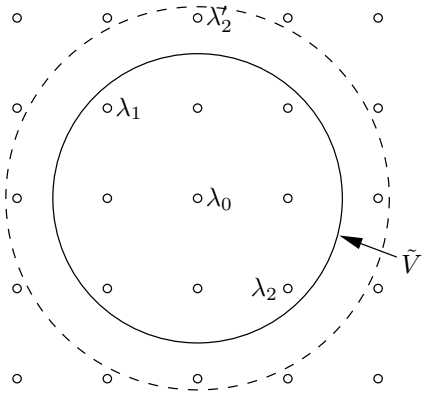}
\end{center}
\caption{The region $\tilde{V}$ is here centered at the point $\lambda_0$. Notice that the distance between $\lambda_1$ and $\lambda_2$ is about twice the maximum distance from $\lambda_0$ to any point in $\Lambda_s\cap \tilde{V}$. The dashed circle illustrates an enlargement of $\tilde{V}$.}
\label{fig:tuples}
\end{figure}

Centering $\tilde{V}$ around $\lambda_0$ points causes a certain asymmetry in the pairwise distances of the elements within a $K$-tuple. Since the region is centered around $\lambda_0$ the maximum pairwise distances between $\lambda_0$ and any other sublattice point will always be smaller than the maximum pairwise distance between any two sublattice points not including $\lambda_0$. 
This can be seen more clearly in Fig.~\ref{fig:tuples}. Notice that the distance between the pair of points labeled $(\lambda_1,\lambda_2)$ is twice the distance than that of the pair $(\lambda_0,\lambda_1)$ or $(\lambda_0,\lambda_2)$. However by slightly increasing the region $\tilde{V}$ to also include $\lambda'_2$ other tuples may be made, which actually have a lower pairwise distance than the pair $(\lambda_1,\lambda_2)$. For this particular example, it is easy to see that the $3$-tuple $t=(\lambda_0,\lambda_1,\lambda_2)$ has a greater SPSD than the $3$-tuple $t'=(\lambda_0,\lambda_1,\lambda'_2)$.

For each $\lambda_0\in V_\pi(0)$ we center a region $\tilde{V}$ around the point, and choose those $N$ $K$-tuples, that give the smallest SPSD\@. By expanding $\tilde{V}$ new $K$-tuples can be constructed that might have a lower SPSD than the SPSD of the original $N$ $K$-tuples. However, the distance from $\lambda_0$ to the points farthest away increases as $\tilde{V}$ increases. Since we only need $N$ $K$-tuples, it can be seen that $\tilde{V}$ should never be larger than twice the lower bound in (\ref{eq:vtilde}) because then the distance from the center to the boundary of the enlarged $\tilde{V}$ region is greater than the maximum distance between any two points in the $\tilde{V}$ region that reaches the lower bound. 
In order to \emph{theoretically} describe the performance of the quantizers, we introduce a dimensionless expansion factor $1\leq \psi_L <2$
which describes how much $\tilde{V}$ must be expanded from the theoretical lower bound~(\ref{eq:vtilde}), to make sure that $N$ optimal $K$-tuples can be constructed by combining sublattice points within a region $\tilde{V}$. 

For the case of $K=2$ we always have $\psi_L=1$ independent of the dimension $L$ so it is only in the case $K\geq 3$ that we need to find expressions for $\psi_L$.
\begin{theorem}\label{theo:psiLK3}
For the case of $K=3$ and any odd $L$ the dimensionless expansion factor is given by 
\begin{equation}
\psi_L=\left(\frac{\omega_L}{\omega_{L-1}}\right)^{1/2L}\left(\frac{L+1}{2L}\right)^{1/2L}\beta_L^{-1/2L},
\end{equation}
where $\omega_L$ is the volume of an $L$-dimensional unit sphere and $\beta_L$ is given by
\begin{equation}\label{eq:betaL}
\begin{split}
\beta_L&=
\sum_{n=0}^{\frac{L+1}2}\binom{\frac{L+1}2}{n}2^{\frac{L+1}2-n}(-1)^n 
\sum_{k=0}^{\frac{L-1}2} \frac{\left(\frac{L+1}2\right)_k \left(\frac{1-L}2\right)_k}{\left(\frac{L+3}2\right)_k\, k!}\\
&\quad \times\sum_{j=0}^k\binom{k}{j}\left(\frac{1}2\right)^{k-j}(-1)^j\left(\frac{1}{4}\right)^j \frac{1}{L+n+j}.
\end{split}
\end{equation}
\end{theorem}
\begin{proof}
See Appendix~\ref{app:theo:psiLK3}
\end{proof}

For the interesting case of $L\rightarrow \infty$ we have the following theorem.
\begin{theorem}\label{theo:psiLinf}
For $K=3$ and $L\rightarrow \infty$ the dimensionless expansion factor $\psi_L$ is given by
\begin{equation}\label{eq:psiinfty}
\psi_{\infty} = \left(\frac{4}{3}\right)^{1/4}.
\end{equation}
\end{theorem}
\begin{proof}
See Appendix~\ref{app:theo:psiLinf}
\end{proof}

Table~\ref{tab:psiL} lists\footnote{Theorem~\ref{theo:psiLK3} is only valid for $L$ odd. However, in the proof of Theorem~\ref{theo:psiLK3} it is straightforward to replace the volume of spherical caps by standard expressions for circle cuts in order to obtain $\psi_2$.} $\psi_L$ for $K=3$ and different values of $L$ and it may be noticed that $\psi_\infty = \sqrt{\psi_1}$. 
\begin{table}[ht]
\begin{center}
\begin{tabular}{c|l}\hline
$L$   & $\psi_L$ \\ \hline
1     & $1.1547005\cdots$ \\
2     & $1.1480804\cdots$ \\
3     & $1.1346009\cdots$ \\
5     & $1.1240543\cdots$ \\
7     & $1.1172933\cdots$ \\
9     & $1.1124896\cdots$ \\
11    & $1.1088540\cdots$ \\
13    & $1.1059819\cdots$ \\ \hline
\end{tabular}\quad
\begin{tabular}{c|l}\hline
$L$   & $\psi_L$ \\ \hline
15    & $1.1036412\cdots$ \\
17    & $1.1016878\cdots$ \\
19    & $1.1000271\cdots$ \\
21    & $1.0985938\cdots$ \\
51    & $1.0883640\cdots$ \\
71    & $1.0855988\cdots$ \\
101   & $1.0831849\cdots$ \\
$\infty$ & $1.0745699\dots$ \\ \hline
\end{tabular}
\end{center}
\caption{$\psi_L$ values obtained by use of Theorems~\ref{theo:psiLK3} and~\ref{theo:psiLinf} for $K=3$.}
\label{tab:psiL}
\end{table}
In order to extend these results to $K>3$ it follows from the proof of Theorem~\ref{theo:psiLK3} that we need closed-form expressions for the volumes of all the different convex regions that can be obtained by $K-1$ overlapping spheres. With such expressions it should be straightforward to find $\psi_L$ for any $K$. However, the analysis of $\psi_L$ for the case of $K=3$ (as given in the proof of Theorem~\ref{theo:psiLK3}) is constructive in the sense that it reveals how $\psi_L$ can be numerically estimated for any $K$ and $L$. Let $\tilde{\nu}$ denote the volume of the expanded sphere $\tilde{V}$. 
Furthermore, let us denote by $T$ the number of $K$-tuples that we construct by using lattice points inside this sphere. Hence, asymptotically as the number of lattice points in $\tilde{V}$ goes to infinity we have
\begin{equation}
T=\left( \frac{\tilde{\nu}/\psi_L^L}{\nu_s}\right)^{K-1},
\end{equation}
which leads to
\begin{equation}
\psi_L = \left(\frac{\omega_L r^L}{\nu_s T^{1/K-1}}\right)^{1/L},
\end{equation}
where $r$ denotes the radius of $\tilde{V}$ and where without loss of generality we can assume that $\nu_s=1$ (simply a matter of scaling). In order to numerically estimate $\psi_L$ it follows that we need to find the set of lattice points within a sphere $\tilde{V}$ of radius $r$. For each of these lattice points we center another sphere of radius $r$ and find the set of lattice points which are within the intersection of the two spheres. This procedure continues $K-1$ times. In the end we find $T$ by adding the number of lattice points within each intersection, i.e.
\begin{equation}
T =
\sum_{\tilde{\Lambda}_1}
\sum_{\tilde{\Lambda}_2}
\dots
\sum_{\tilde{\Lambda}_{K-2}}
|\Lambda_s \cap \tilde{V}(\lambda_{K-2})\cap \dots \cap \tilde{V}(\lambda_0)|,
\end{equation}
where
\begin{equation}
\begin{split}
\tilde{\Lambda}_1 &= \{ \lambda_1 : \lambda_1\in \Lambda_s\cap \tilde{V}(\lambda_0) \}, \\
\tilde{\Lambda}_2 &= \{ \lambda_2 : \lambda_2\in \Lambda_s\cap \tilde{V}(\lambda_1)\cap \tilde{V}(\lambda_0)\}, \\
&\hspace{2mm}\vdots \\
\tilde{\Lambda}_{K-2} &= \{\lambda_{K-2} : \lambda_{K-2} \in \Lambda_s \cap
\tilde{V}(\lambda_{K-3})\cap \dots \cap \tilde{V}(\lambda_0) \}.
\end{split}
\end{equation}

For example for $K=4, \Lambda=Z^2$ and $r=10,20,50$ and $70$ then using the algorithm outlined above we find $\psi_2\approx 1.1672, 1.1736, 1.1757$ and $1.1762$, respectively.
\begin{remark}
In order to achieve the shift-invariance property of the index-assignment algorithm, we impose a restriction upon $\lambda_0$ points. Specifically, we require that $\lambda_0\in V_\pi(0)$ so that the first coordinate of any $K$-tuple is within the region $V_\pi(0)$. To avoid excluding $K$-tuples that have their first coordinate outside $V_\pi(0)$ we form cosets of each $K$-tuple and allow only one member from each coset to be assigned to a central lattice point within $V_\pi(0)$. This restriction, which is only put on $\lambda_0\in \Lambda_s$ , might cause a bias towards $\lambda_0$ points. 
However, it is easy to show that, asymptotically as $N\rightarrow \infty$, any such bias can be removed. For the case of $K=2$ we can use similar arguments as used in~\cite{diggavi:2002} and for $K>2$ we can show that the amount of $K$-tuples that is affected by this restriction is small compared to the amount of $K$-tuples which are not affected. 
Hence, asymptotically as $N\rightarrow \infty$, this restriction is effectively removed. So for example this means that we can enforce similar restriction on all sublattice points, which, asymptotically as $N\rightarrow\infty$, will only reduce the number of $K$-tuples by a neglectable amount. And as such, any possible bias towards the set of points $\lambda_0\in\Lambda_s$ is removed. 

As mentioned above, the $K$-tuples need to be assigned to central lattice points within $V_\pi(0)$. This is a standard linear assignment problem where a cost measure is minimized. However, solutions to linear assignment problems are generally not unique. Therefore, there might exist several labelings, which all yield the same cost, but exhibit a different amount of asymmetry. Theoretically, exact symmetry may then be obtained by e.g.\ time-sharing through a suitable mixing of labelings. 
In practice, however, any scheme would use a finite $N$ (and finite rates). In addition, for many applications, time-sharing is inconvenient. In these non-asymptotical cases we cannot guarantee exact symmetry. To this end, we have provided a few examples that assess the distortions obtained from practical experiments, see Section~\ref{sec:num} (Tables~\ref{tab:kappa1} and~\ref{tab:kappa2}).
\end{remark}

\section{High-Resolution Analysis}\label{sec:highrate}
In this section we derive high-resolution approximations for the expected distortion. For this high-resolution analysis we let $N\rightarrow \infty$ and $\nu_s\rightarrow 0$. The effect of this is that the index of sublattice increases, but the actual volumes of the Voronoi cells shrink.

\subsection{Total Expected Distortion}
We first introduce Conjecture~\ref{con:riemann2} which relates the sum of distances between pairs of sublattice points to $G(S_L)$, the dimensionless normalized second moment of an $L$-dimensional sphere. In Appendix~\ref{app:riemann2} we prove the conjecture for the case of $K=2$ and any $L$ as well as for the case of $K=3$ and $L\rightarrow \infty$. In addition we show in Appendix~\ref{app:riemann2} that Conjecture~\ref{con:riemann2} is a good approximation for the case of $K=3$ and finite $L$. After presenting Conjecture~\ref{con:riemann2} we determine the dominating term in the expression for the expected distortion. This is given by Proposition~\ref{prop:growthriemann2}.

\begin{conjecture}\label{con:riemann2}
For $L,N\rightarrow \infty$ and $\nu_s\rightarrow 0$, we have
for any pair $(i,j),\ i,j=0,\dots,K-1,\ i\neq j$,
\begin{equation*}
\sum_{\lambda_c\in V_\pi(0)}\!\!\!\| \alpha_i(\lambda_c)-\alpha_j(\lambda_c)\|^2
= G(S_L)\psi_L^{2}N^2N^{2K/L(K-1)}\nu^{2/L}.
\end{equation*}
\end{conjecture}

\begin{proposition}\label{prop:growthriemann2}
For $N\rightarrow \infty$ and $2\leq K<\infty$ we have
\begin{equation}
\mathcal{O}\left(
\frac{\sum_{\lambda_c\in V_\pi(0)}\left\| \lambda_c - \frac{1}{K}\sum_{i=0}^{K-1}\lambda_i\right\|^2}
{\sum_{\lambda_c\in V_\pi(0)}\sum_{i=0}^{K-2}\sum_{j=i+1}^{K-1}\| \lambda_i - \lambda_j\|^2}\right) \rightarrow 0.
\end{equation}
\end{proposition}

\begin{proof}
See Appendix~\ref{app:growthriemann2}.
\end{proof}

The expected distortion (\ref{eq:expdist}) can by use of Theorem~\ref{theo:sums} be written as
\begin{equation}\label{eq:da}
\begin{split}
&d_a^{(K,\kappa)}\approx (1-p)^{\kappa}p^{K-\kappa} \\
&\quad \times\left(
\binom{K}{\kappa}d_c 
+ \frac{1}{N^2}
\sum_{l\in \mathcal{L}}\sum_{\lambda_c\in V_\pi(0)}
\left\|\lambda_c - \frac{1}{\kappa}\sum_{j=0}^{\kappa-1} \lambda_{l_j}\right\|^2\right) \\ 
&=
(1-p)^{\kappa}p^{K-\kappa}\binom{K}{\kappa}\\
&\quad\times\Bigg(d_c +
\frac{1}{N^2}\sum_{\lambda_c\in V_\pi(0)}
\left(
\left\| \lambda_c - \frac{1}{K}\sum_{i=0}^{K-1}\lambda_i\right\|^2
\right.
\\
&\quad+\left.
\left(\frac{K-\kappa}{K^2\kappa(K-1)}\right)
\sum_{i=0}^{K-2}\sum_{j=i+1}^{K-1}\| \lambda_i - \lambda_j\|^2 
\right)\Bigg).
\end{split}
\end{equation}

By use of Conjecture~\ref{con:riemann2} (as an approximation that becomes exact for $L\rightarrow \infty$), Proposition~\ref{prop:growthriemann2} and Eq.~(\ref{eq:d0G}) it follows that~(\ref{eq:da}) can be written as
\begin{align}\notag
&d_a^{(K,\kappa)}\approx (1-p)^{\kappa}p^{K-\kappa}\binom{K}{\kappa}\\ \notag
&\times\!\!\left(\!d_c +
\frac{1}{N^2}\!\!\sum_{\lambda_c\in V_\pi(0)}\!\!
\left(\frac{K-\kappa}{K^2\kappa(K-1)}\right)
\sum_{i=0}^{K-2}\sum_{j=i+1}^{K-1}\| \lambda_i - \lambda_j\|^2 
\right)\\ \notag 
&\approx (1-p)^{\kappa}p^{K-\kappa}\binom{K}{\kappa} \\ \label{eq:distkleqK}
&\times\left(G(\Lambda_c)\nu^{2/L} +
\left(\frac{K-\kappa}{2K\kappa}\right)
G(S_L)\psi_L^{2}N^{2K/L(K-1)}\nu^{2/L}
\right).
\end{align}
The second term in~(\ref{eq:distkleqK}) is the dominating term for $\kappa<K$ and $N\rightarrow \infty$. Observe\footnote{This was pointed out by a reviewer who also drew the connection to recent results based on source-channel erasure codes~\cite{pradhan:2004} where the improvement by receiving more descriptions is almost linear in certain cases.} that this term is only dependent upon $\kappa$ through the coefficient $\frac{K-\kappa}{2K\kappa}$.

The total expected distortion is obtained by summing over $\kappa$ including the cases where $\kappa=0$ and $\kappa=K$,
\begin{equation}\label{eq:adistopt}
\begin{split}
d_a &\approx \hat{K}_1G(\Lambda_c)\nu^{2/L} +\hat{K}_2 
G(S_L)\psi_L^{2}N^{2K/L(K-1)}\nu^{2/L}\\
&\quad + p^K E[\|X\|^2],
\end{split}
\end{equation}
where $\hat{K}_1$ is given by
\begin{equation}
\begin{split}
\hat{K}_1&=\sum_{\kappa=1}^K\binom{K}{\kappa}p^{K-\kappa}(1-p)^\kappa\\
&= 1-p^K.
\end{split}
\end{equation}
and $\hat{K}_2$ is given by
\begin{equation}\label{eq:K2}
\hat{K}_2=\sum_{\kappa=1}^K\binom{K}{\kappa}p^{K-\kappa}(1-p)^\kappa\frac{K-\kappa}{2\kappa K}.
\end{equation}

Using~(\ref{eq:R0}) and~(\ref{eq:Rs}) we can write $\nu$ and $N$ as a function of differential entropy and side entropies, that is
\begin{equation}
\nu^{2/L}=2^{2(h(X)-R_c)},
\end{equation}
and
\begin{equation}
N^{2K/L(K-1)}=2^{\frac{2K}{K-1}(R_c-R_s)},
\end{equation}
from which we may write the expected distortion as a function of entropies, that is
\begin{equation}\label{eq:daopt}
\begin{split}
d_a &\approx \hat{K}_1G(\Lambda_c)2^{2(h(X)-R_c)}\\
&\quad+ \hat{K}_2\psi_L^{2}G(S_L) 2^{2(h(X)-R_c)}2^{\frac{2K}{K-1}(R_c-R_s)} 
+ p^K E[\|X\|^2],
\end{split}
\end{equation}
where we see that the distortion due to the side quantizers only depends upon the scaling (and dimension) of the sublattice and not which sublattice is used.

\subsection{Optimal $\nu$, $N$ and $K$.}\label{sec:opt}
We now derive expressions for the optimal $\nu$, $N$ and $K$. Using these values we are able to construct the lattices $\Lambda_c$ and $\Lambda_s$. The optimal index assignment is hereafter found by using the approach outlined in Section~\ref{sec:label}. These lattices combined with their index assignment completely specify an optimal entropy-constrained MD-LVQ system.

In order for the entropies of the side descriptions to be equal to the target entropy $R_t/K$, we rewrite (\ref{eq:Rs}) and get
\begin{equation}\label{eq:tau}
N\nu = 2^{L(h(X)-R_t/K)} \triangleq \tau,
\end{equation}
where $\tau$ is constant. 
The expected distortion may now be expressed as a function of $\nu$,
\begin{equation}
\begin{split}
d_a&=\hat{K}_1G(\Lambda_c)\nu^{2/L}\\  
&+\hat{K}_2\psi_L^{2}G(S_L) \nu^{2/L}\nu^{-\frac{2K}{L(K-1)}}\tau^{-\frac{2K}{L(K-1)}}+ p^K E[\|X\|^2].
\end{split}
\end{equation}
Differentiating w.r.t.\ $\nu$ and equating to zero gives,
\begin{equation}\label{eq:diffda}
\begin{split}
&0=\frac{\partial d_a}{\partial \nu}=
\frac{2}{L}\hat{K}_1G(\Lambda_c)\frac{\nu^{2/L}}{\nu}\\
&+ \left(\frac{2}{L} - \frac{2}{L}\frac{K}{K-1}\right)\hat{K}_2\psi_L^{2}G(S_L) \frac{\nu^{2/L}}{\nu}\nu^{-\frac{2K}{L(K-1)}}\tau^{-\frac{2K}{L(K-1)}},
\end{split}
\end{equation}
from which we obtain the optimal value of $\nu$
\begin{equation}\label{eq:optnu}
\nu = \tau\left(\frac{1}{K-1} \frac{\hat{K}_2}{\hat{K}_1} \frac{G(S_L)}{G(\Lambda_c)}\psi_L^{2}\right)^{\frac{L(K-1)}{2K}}.
\end{equation}
The optimal $N$ follows easily by use of (\ref{eq:tau})
\begin{equation}\label{eq:optN}
N = \left((K-1)\frac{\hat{K}_1}{\hat{K}_2} \frac{G(\Lambda_c)}{G(S_L)}\frac{1}{\psi_L^{2}}\right)^{\frac{L(K-1)}{2K}}.
\end{equation}
Eq.~(\ref{eq:optN}) shows that the optimal redundancy $N$ is, for a fixed $K$, independent of the sublattice as well as the target entropy.

For a fixed $K$ the optimal $\nu$ and $N$ are given by~(\ref{eq:optnu}) and~(\ref{eq:optN}), respectively, and the optimal $K$ can then easily be found by evaluating~(\ref{eq:adistopt}) for various values of $K$, and choosing the one that yields the lowest expected distortion. The optimal $K$ is then given by
\begin{equation}\label{eq:optK}
K_\text{opt}=\arg\,\min_{K} d_a,\quad K=1,\dots, K_\text{max},
\end{equation}
where $K_\text{max}$ is a suitable chosen positive integer. In practice $K$ will always be finite and furthermore limited to a narrow range of integers, which makes the complexity of the minimization approach, given by~(\ref{eq:optK}), negligible.

\section{Construction of Quantizers}
\label{sec:construction}
In this section we design practical quantizers. We show that the index values are restricted to a discrete set of admissible values. Knowledge of these values makes it possible to construct practical quantizers and theoretically describe their performance.

\subsection{Index Values}
Eqs.\ (\ref{eq:optnu}) and (\ref{eq:optN}) suggest that we are able to continuously trade-off central versus side-distortions by adjusting $N$ and $\nu$ according to the packet-loss probability. 
This is, however,  not the case, since certain constraints must be imposed on $N$. 
First of all, since $N$ denotes the number of central lattice points within each Voronoi cell of the sublattice, it must be integer and positive. 
Second, we require the sublattice to be geometrical similar to the central lattice. Finally, we require the sublattice to be a clean sublattice, so that no central lattice points are located on boundaries of Voronoi cells of the sublattice. This restrict the amount of admissible index values for a particular lattice to a discrete set, c.f.~\cite{diggavi:2002}.

\begin{figure}[ht]
\begin{center}
\includegraphics[width=8cm]{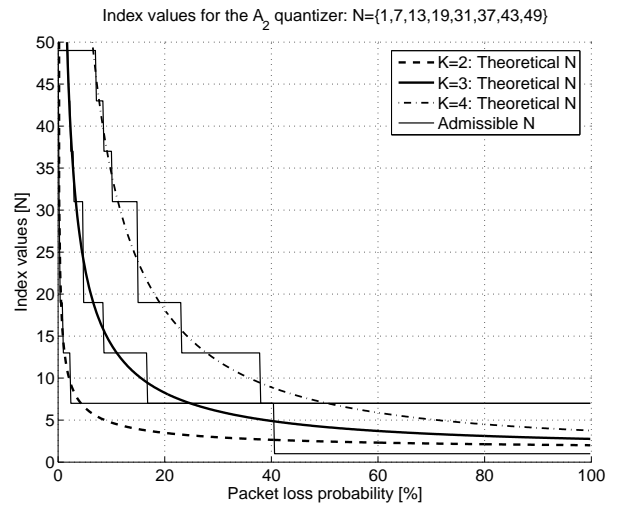}
\caption{Theoretical optimal index values for the $A_2$ quantizer as a function of packet-loss probability. Thin solid lines are obtained by restricting the theoretical optimal index values given by (\ref{eq:optN}) to optimal admissible values. The optimal admissible index values are those that minimize (\ref{eq:adistopt}) for a given $p$.}
\label{fig:index_values}
\end{center}\end{figure}

Fig.~\ref{fig:index_values} shows the theoretically optimal index values (i.e.\ ignoring the fact that $N$ belongs to a discrete set) for the $A_2$ quantizer, given by (\ref{eq:optN}) for $\psi_L=1,1.1481$ and $1.1762$ corresponding to $K=2,3$ and $4$, respectively. Also shown are the theoretical optimal index values when restricted to admissible index values. 
Notice that the optimal index value $N$ increases for increasing number of descriptions. 
This is to be expected since a higher index value leads to less redundancy;
this redundancy reduction, however, is balanced out by the redundancy increase resulting from the added number of descriptions.
In \cite{ostergaard:2004} we observed that for a two-description system, usually only very few index values would be used. In fact for the two-dimensional $A_2$ quantizer, only $N\in\{1,7,13\}$ should be used. Higher dimensional quantizers would use greater index values. However, here we see that by increasing the number of descriptions beyond $K=2$, it is optimal to use greater index values which adds more flexibility to the scheme.

From Fig.~\ref{fig:index_values} it can be seen that when the continuous optimal index value is rounded to the optimal admissible index value it is always the closest one from either below or above. This means that the optimal admissible index value is found by considering only the two values closest to the continuous index value, and using the one that minimizes (\ref{eq:adistopt}).

\section{Numerical Evaluation}\label{sec:num}
In this section we compare the numerical performances of two-dimensional entropy-constrained MD-LVQ systems (based on the $A_2$ lattice) to their theoretical prescribed performances. 

\subsection{Performance of Individual Descriptions}
In the first experiment we design a 3-channel MD-LVQ based on the $A_2$ quantizer. We quantize an i.i.d.\ unit-variance zero-mean Gaussian source which has been blocked into two-dimensional vectors. The number of vectors used in the experiment is $2\cdot10^6$. The entropy of each side description is 5 bit/dim.\ and we vary the index value in the range $31$ -- $67$. The dimensionless expansion factor $\psi_L$ is set to $1.14808$. The numerical and theoretical distortions when receiving only a single description out of the three is shown in Table~\ref{tab:kappa1}. Similarly, Table~\ref{tab:kappa2} shows the distortions of the same system due to reception of two out of three descriptions and Table~\ref{tab:kappa3} shows the performance of the central quantizer when all three descriptions are received. The column labeled ``Avg.'' illustrates the average distortion of the three numerically measured distortions and the column labeled ``Theo.'' describes the theoretical distortions given by~(\ref{eq:distkleqK})~\footnote{Since we do not consider packet-losses in this experiment we have set the weight to unity, i.e.\ $(1-p)^\kappa p^{K-1}\binom{K}{\kappa}=1$.}. It is clear from the tables that the system is symmetric; the achieved distortion depends on the number of received descriptions but is essentially independent of \emph{which} descriptions are used for reconstruction.
\begin{table}[ht]
\begin{center}
\begin{tabular}{cccccc}
$N$ & $\lambda_0$ &  $\lambda_1$ &  $\lambda_2$ & Avg. & Theo. \\ \hline
31 & $-25.6918$ & $-25.6875$ & $-25.6395$ & $-25.6729$ & $-24.8280$ \\
37 & $-24.5835$ & $-24.5324$ & $-24.5404$ & $-24.5521$ & $-24.4571$ \\
43 & $-24.5772$ & $-24.5972$ & $-24.5196$ & $-24.5647$ & $-24.1396$ \\
49 & $-24.2007$ & $-24.2837$ & $-24.2713$ & $-24.2519$ & $-23.8622$ \\
61 & $-23.8616$ & $-23.9011$ & $-23.8643$ & $-23.8757$ & $-23.3946$ \\
67 & $-23.7368$ & $-23.7362$ & $-23.7655$ & $-23.7462$ & $-23.1936$ \\ \hline
\end{tabular}
\caption{Distortion [dB] due to reception of a single description out of three.}
\label{tab:kappa1}
\end{center}
\end{table}
\begin{table}[ht]
\begin{center}
\begin{tabular}{p{1mm}p{12mm}p{12mm}p{12mm}cc}
$N$ & \mbox{$\frac{1}2(\lambda_0+\lambda_1)$} & \mbox{$\frac{1}2(\lambda_0+\lambda_2)$} & \mbox{$\frac{1}2(\lambda_1+\lambda_2)$} & Avg. & Theo. \\ \hline
31 & $-30.7792$ & $-30.7090$ & $-30.7123$ & $-30.7335$ & $-30.6810$ \\
37 & $-29.8648$ & $-29.8430$ & $-29.9472$ & $-29.8850$ & $-30.3482$ \\
43 & $-29.9087$ & $-29.8749$ & $-29.9641$ & $-29.9159$ & $-30.0563$ \\
49 & $-29.6290$ & $-29.5577$ & $-29.6662$ & $-29.6176$ & $-29.7971$ \\
61 & $-29.3076$ & $-29.2185$ & $-29.3715$ & $-29.2992$ & $-29.3532$ \\
67 & $-29.1752$ & $-29.2128$ & $-29.2151$ & $-29.2010$ & $-29.1603$ \\ \hline
\end{tabular}
\caption{Distortion [dB] due to reception of two descriptions out of three.}
\label{tab:kappa2}
\end{center}
\end{table}

\begin{table}[ht]
\begin{center}
\begin{tabular}{ccc}
$N$ & $\lambda_c$ & Theo. \\ \hline
31  & $-43.6509$ & $-43.6508$ \\ 
37  & $-44.4199$ & $-44.4192$ \\
43  & $-45.0705$ & $-45.0719$ \\
49  & $-45.6401$ & $-45.6391$ \\
61  & $-46.5879$ & $-46.5905$ \\
67  & $-46.9992$ & $-46.9979$ \\ \hline
\end{tabular}
\caption{Distortion [dB] due to reception of all three descriptions out of three.}
\label{tab:kappa3}
\end{center}
\end{table}

\subsection{Distortion as a Function of Packet-Loss Probability}
\begin{figure}[ht]
\begin{center}
\includegraphics[width=8cm]{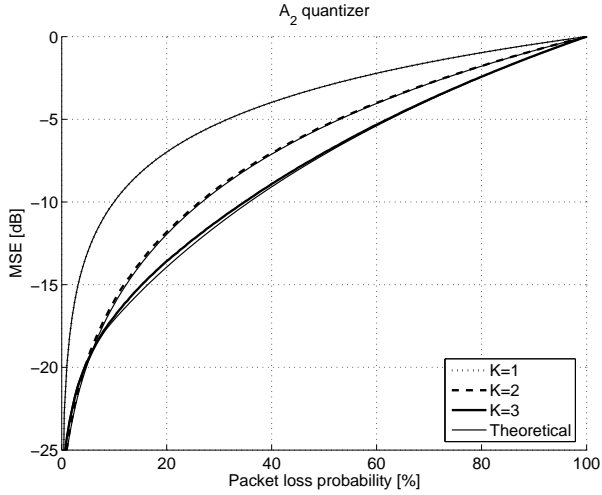}
\caption{Distortion as a function of packet-loss probability for the $A_2$ quantizer. The target entropy is 6 bit/dim., so each description gets 6/K bit/dim. Thick lines show numerical performance and thin solid lines show theoretical performance.}
\label{fig:numdistpp}
\vspace{-1cm}
\end{center}\end{figure}

We now show the expected distortion as a function of the packet-loss probability for $K$-channel MD-LVQ systems where $K=1,2,3$. We block the i.i.d.\ unit-variance Gaussian source into $2\cdot 10^6$ two-dimensional vectors and let the total target entropy be 6 bit/dim. The expansion factor is set to $\psi_2=1$ for $K=1,2$ and $\psi_2=1.14808$ for $K=3$. We sweep the packet-loss probability $p$ in the range $p\in [0;1]$ in steps of 1/200 and for each $p$ we measure the distortion for all admissible index values and use that index value which gives the lowest distortion. This gives rise to an operational lower hull (OLH) for each quantizer. This is done for the theoretical curves as well by inserting admissible index values in~(\ref{eq:adistopt}) and use that index value that gives the lowest distortion. 
In other words we compare the numerical OLH with the theoretical OLH and not the ``true''\footnote{A lattice is restricted to a set of admissible index values. This set is generally expanded when the lattice is used as a product quantizer, hence admissible index values closer to the optimal values given by~(\ref{eq:optN}) can in theory be obtained.} lower hull that would be obtained by using the unrestricted index values given by (\ref{eq:optN}).
The target entropy is evenly distributed over $K$ descriptions. For example, for $K=2$ each description uses 3 bit/dim., whereas for $K=3$ each description uses only 2 bit/dim. The performance is shown shown in Fig.~\ref{fig:numdistpp}. The practical performance of the scheme is described by the lower hull of the $K$-curves. Notice that at higher packet-loss probabilities ($p>5\%$) it becomes advantageous to use three descriptions instead two. 
%
%

\section{Conclusion and Discussion}
In this work we derived analytical expressions for the central and side quantizers which, under high-resolutions assumptions, minimize the expected distortion of a symmetric n-channel MD-LVQ subject to entropy constraints on the side descriptions for given packet-loss probabilities. 
The expected distortion observed at the receiving side depends only upon the number of received descriptions but is independent of which descriptions are received. 
We focused on a special case of the symmetric multiple-description problem where only a single parameter controls the redundancy tradeoffs between the central and the side distortions. As such more work is needed before the general symmetric $n$-channel MD-LVQ problem is completely solved. A step in that direction is presented in~\cite{ostergaard:2006b}.

Future work in progress includes extending the presented scheme to the asymmetric case, where packet-loss probabilities, entropies and distortions may differ for the different descriptions~\cite{ostergaard:2005d}.


%
%
\appendices
\section{Proof of Theorem~\ref{theo:sums}}
\label{app:sums}
In order to prove Theorem~\ref{theo:sums}, we need the following results.

\begin{lemma}\label{lem:r2}
For $1\leq \kappa \leq K$ we have
\begin{equation*}
\begin{split}
\sum_{l\in\mathcal{L}}
\left\langle \lambda_c,\sum_{j=0}^{\kappa-1} \lambda_{l_j}\right\rangle 
&= 
\frac{\kappa}{K}\binom{K}{\kappa}
\left\langle \lambda_c, \sum_{i=0}^{K-1} \lambda_i\right\rangle.
\end{split}
\end{equation*}
\end{lemma}

\begin{proof}
Expanding all sums on the left-hand-side leads to $\binom{K}{\kappa}\kappa$ different terms of the form $\langle \lambda_c, \lambda_i\rangle$, where $i\in \{0,\dots, K-1\}$. There are $K$ distinct $\lambda_i$'s so the number of times each $\lambda_i$ occur is $\binom{K}{\kappa}\kappa/K$.
\end{proof}

\begin{lemma}\label{lem:r1}
For $1\leq \kappa \leq K$ we have
\begin{equation*}
\begin{split}
&\sum_{l\in\mathcal{L}}\left\| \sum_{j=0}^{\kappa-1} \lambda_{l_j}\right\|^2\\
&=
\frac{\kappa}{K}\binom{K}{\kappa}\sum_{i=0}^{K-1}\|\lambda_i\|^2
+\frac{2\kappa(\kappa-1)}{K(K-1)}\binom{K}{\kappa}
\sum_{i=0}^{K-2}\sum_{j=i+1}^{K-1}\langle \lambda_i,\lambda_j \rangle.
\end{split}
\end{equation*}
\end{lemma}

\begin{proof}
There are $\binom{K}{\kappa}$ distinct ways of adding $\kappa$ out of $K$ elements. Squaring a sum of $\kappa$ elements leads to $\kappa$ squared elements and $2\binom{\kappa}{2}$ cross products (product of two different elements). This gives a total of $\binom{K}{\kappa}\kappa$ squared elements, and $2\binom{K}{\kappa}\binom{\kappa}{2}$ cross products. Now since there are $K$ distinct elements, the number of times each squared element occurs is given by
\begin{equation}\label{eq:lemnum1}
\#_{\|\lambda_i\|^2} =\binom{K}{k}\frac{\kappa}{K}.
\end{equation}
 There are $\binom{K}{2}$ distinct cross products, so the number of times each cross product occurs is given by
\begin{equation}\label{eq:lemnum2}
\#_{\langle\lambda_i,\lambda_j\rangle} = 
\binom{K}{\kappa}\frac{2\binom{\kappa}{2}}{\binom{K}{2}}=
\displaystyle\frac{2\kappa(\kappa-1)}{K(K-1)}\displaystyle\binom{K}{\kappa}.
\end{equation}
\rspace
\end{proof}

\begin{lemma}\label{lem:r3}
For $K\geq 1$ we have
\begin{equation}\label{eq:r3}
\begin{split}
(K-1)\sum_{i=0}^{K-1}\|\lambda_i\|^2 &- 2\sum_{i=0}^{K-2}\sum_{j=i+1}^{K-1}\langle \lambda_i , \lambda_j \rangle\\
&= \sum_{i=0}^{K-2}\sum_{j=i+1}^{K-1} \|\lambda_i - \lambda_j\|^2.
\end{split}
\end{equation}
\end{lemma}
\begin{proof}
Expanding the right-hand-side of~(\ref{eq:r3}) yields
\begin{equation}
\begin{split}
\sum_{i=0}^{K-2}\sum_{j=i+1}^{K-1}&\|\lambda_i - \lambda_j\|^2\\ 
&=\sum_{i=0}^{K-2}\sum_{j=i+1}^{K-1}\left( \|\lambda_i\|^2 + \|\lambda_j\|^2 -
2\langle \lambda_i, \lambda_j \rangle \right).
\end{split}
\end{equation}
We also have
\begin{equation}
\begin{split}
\sum_{i=0}^{K-2}&\sum_{j=i+1}^{K-1}\left( \|\lambda_i\|^2 + \|\lambda_j\|^2\right)\\
&= 
\sum_{i=0}^{K-2}(K-1-i)\|\lambda_i\|^2 + \sum_{i=0}^{K-2}\sum_{j=i+1}^{K-1}\|\lambda_j\|^2 \\
&= \sum_{i=0}^{K-2}(K-1-i)\|\lambda_i\|^2 + \sum_{j=1}^{K-1}j\|\lambda_j\|^2 \\
&= \sum_{i=0}^{K-1}(K-1-i)\|\lambda_i\|^2 + \sum_{j=0}^{K-1}j\|\lambda_j\|^2 \\
&= \sum_{i=0}^{K-1}(K-1)\|\lambda_i\|^2 - \sum_{i=0}^{K-1}i\|\lambda_i\|^2 + \sum_{j=0}^{K-1}j\|\lambda_j\|^2 \\
&= (K-1)\sum_{i=0}^{K-1}\|\lambda_i\|^2,
\end{split}
\end{equation}
which completes the proof.
\end{proof}

We are now in a position to prove the following result.
\begin{proposition}{\label{prop:sums}}
For $1\leq \kappa \leq K$ we have
\begin{equation*}
\begin{split}
&\sum_{l\in\mathcal{L}}\left\| \lambda_c - \frac{1}{\kappa}
\sum_{j=0}^{\kappa-1} \lambda_{l_j}\right\|^2= 
\binom{K}{\kappa}
\Bigg(
\left\| \lambda_c - 
\frac{1}{K}\sum_{i=0}^{K-1}\lambda_i\right\|^2
\\
&\quad+\left(
\frac{K-\kappa}{K^2\kappa(K-1)}\right)
\sum_{i=0}^{K-2}\sum_{j=i+1}^{K-1}\| \lambda_i - \lambda_j\|^2
\Bigg).
\end{split}
\end{equation*}
\end{proposition}

\begin{proof}
We have
\begin{equation*}\label{eq:costfunctional4}
\begin{split}
\left\| \lambda_c - \frac{1}{\kappa}\sum_{j=0}^{\kappa-1} \lambda_{l_j}\right\|^2
&=\|\lambda_c\|^2- 2\left\langle \lambda_c,\frac{1}{\kappa}\sum_{j=0}^{\kappa-1} \lambda_{l_j}\right\rangle \\
&\quad+
\frac{1}{\kappa^2}\left\| \sum_{j=0}^{\kappa-1} \lambda_{l_j}\right\|^2.
\end{split}
\end{equation*}
Hence, by use of Lemmas~\ref{lem:r2} and~\ref{lem:r1}, we have that
\begin{equation*}
\begin{split}
&\sum_{l\in\mathcal{L}}
\left\| \lambda_c - \frac{1}{\kappa}\sum_{j=0}^{\kappa-1} \lambda_{l_j}\right\|^2 \\
&=
\binom{K}{\kappa}\Bigg(
\|\lambda_c\|^2 
- \frac{2}{K}\left\langle \lambda_c,\sum_{i=0}^{K-1}\lambda_i\right\rangle
+
\frac{1}{K\kappa}\sum_{i=0}^{K-1}\|\lambda_i\|^2 \\
&\quad+\frac{2(\kappa-1)}{K(K-1)\kappa}
\sum_{i=0}^{K-2}\sum_{j=i+1}^{K-1}\langle \lambda_i,\lambda_j \rangle
\Bigg) \\
&=
\binom{K}{\kappa}\Bigg(
\left\| \lambda_c - 
\frac{1}{K}\sum_{i=0}^{K-1}\lambda_i\right\|^2
-
\frac{1}{K^2}
\left\|\sum_{i=0}^{K-1}\lambda_i\right\|^2\\
&\quad+
\frac{1}{K\kappa}\sum_{i=0}^{K-1}\|\lambda_i\|^2
+\frac{2(\kappa-1)}{K(K-1)\kappa}
\sum_{i=0}^{K-2}\sum_{j=i+1}^{K-1}\langle \lambda_i,\lambda_j \rangle
\Bigg) \\
&=
\binom{K}{\kappa}\Bigg(
\left\| \lambda_c - 
\frac{1}{K}\sum_{i=0}^{K-1}\lambda_i\right\|^2
+
\left(\frac{1}{K\kappa} - \frac{1}{K^2}\right)
\sum_{i=0}^{K-1}\|\lambda_i\|^2\\
&\quad+
\left(\frac{2(\kappa-1)}{K(K-1)\kappa}-\frac{2}{K^2}\right)
\sum_{i=0}^{K-2}\sum_{j=i+1}^{K-1}\langle \lambda_i,\lambda_j \rangle
\Bigg) \\
&=
\binom{K}{\kappa}\Bigg(
\left\| \lambda_c - 
\frac{1}{K}\sum_{i=0}^{K-1}\lambda_i\right\|^2
+
\left(\frac{K-\kappa}{K^2\kappa}\right)
\sum_{i=0}^{K-1}\|\lambda_i\|^2 \\
&\quad-
\left(\frac{K-\kappa}{K^2\kappa(K-1)}\right)
2\sum_{i=0}^{K-2}\sum_{j=i+1}^{K-1}\langle \lambda_i,\lambda_j \rangle
\Bigg)
\intertext{so that, by Lemma~\ref{lem:r3}, we finally have that}
&=
\binom{K}{\kappa}\Bigg(
\left\| \lambda_c - 
\frac{1}{K}\sum_{i=0}^{K-1}\lambda_i\right\|^2 \\
&\quad+
\left(\frac{K-\kappa}{K^2\kappa(K-1)}\right)
\sum_{i=0}^{K-2}\sum_{j=i+1}^{K-1}\| \lambda_i-\lambda_j \|^2
\Bigg),
\end{split}
\end{equation*}
which completes the proof.
\end{proof}

\begin{theo_empty}{\ref{theo:sums}.}
For $1\leq \kappa \leq K$ we have
\begin{equation*}
\begin{split}
&\sum_{\lambda_c}\sum_{l\in\mathcal{L}}
\left\| \lambda_c - \frac{1}{\kappa}\sum_{j=0}^{\kappa-1} \lambda_{l_j}\right\|^2
= \sum_{\lambda_c}\binom{K}{\kappa}\Bigg(\left\| \lambda_c - 
\frac{1}{K}\sum_{i=0}^{K-1}\lambda_i\right\|^2\\
&\quad+
\left(
\frac{K-\kappa}{K^2\kappa(K-1)}\right)
\sum_{i=0}^{K-2}\sum_{j=i+1}^{K-1}\| \lambda_i - \lambda_j\|^2\Bigg).
\end{split}
\end{equation*}
\end{theo_empty}

\begin{proof}
Follows trivially from Proposition~\ref{prop:sums}.
\end{proof}

\section{Proof of Theorem~\ref{theo:psiLK3}}\label{app:theo:psiLK3}
\begin{theo_empty}{\ref{theo:psiLK3}.}
For the case of $K=3$ and any odd $L$ the dimensionless expansion factor is given by 
\begin{equation}
\psi_L=\left(\frac{\omega_L}{\omega_{L-1}}\right)^{1/2L}\left(\frac{L+1}{2L}\right)^{1/2L}\beta_L^{-1/2L},
\end{equation}
where $\beta_L$ is given by
\begin{equation}\label{eq:betaL1}
\begin{split}
\beta_L&=
\sum_{n=0}^{\frac{L+1}2}\binom{\frac{L+1}2}{n}2^{\frac{L+1}2-n}(-1)^n 
\sum_{k=0}^{\frac{L-1}2} \frac{\left(\frac{L+1}2\right)_k \left(\frac{1-L}2\right)_k}{\left(\frac{L+3}2\right)_k\, k!}\\
&\quad\times\sum_{j=0}^k\binom{k}{j}\left(\frac{1}2\right)^{k-j}(-1)^j\left(\frac{1}{4}\right)^j \frac{1}{L+n+j}.
\end{split}
\end{equation}
\end{theo_empty}

\begin{proof}
In the following we consider the case of $K=3$. For a specific $\lambda_0\in \Lambda_s$ we need to construct $N$ $3$-tuples all having $\lambda_0$ as the first coordinate. To do this we first center a sphere $\tilde{V}$ of radius $r$ at $\lambda_0$. For large $N$ and small $\nu_s$ this sphere contains approximately $\tilde{\nu}/\nu_s$ lattice points from $\Lambda_s$. Hence, it is possible to construct $(\tilde{\nu}/\nu_s)^2$ distinct 3-tuples. However, the maximum distance between $\lambda_1$ and $\lambda_2$ points is greater than the maximum distance between $\lambda_0$ and $\lambda_1$ points and also between $\lambda_0$ and $\lambda_2$ points. To avoid this bias towards $\lambda_0$ points we make sure that we only use 3-tuples that satisfy $\|\lambda_i-\lambda_j\|\leq r/\sqrt{L}$ for $i,j=0,1,2$. However, with this restriction we can no longer form $N$ 3-tuples. Therefore, we expand $\tilde{V}$ by the factor $\psi_L$ in order to make sure that exactly $N$ 3-tuples can be made. 
It is well known that the number of lattice points at exactly squared distance $l$ from $c$, for any $c \in \mathbb{R}^L$ is given by the coefficients of the Theta series of the lattice $\Lambda$~\cite{conway:1999}. Theta series depend on the lattices and also on $c$~\cite{conway:1999}. Instead of working directly with Theta series we will, in order to be lattice and displacement independent, consider the $L$-dimensional \emph{hollow} sphere $\bar{\mathcal{C}}$ obtained as $\bar{\mathcal{C}}=S(c,m)-S(c,m-1)$ and shown in Fig.~\ref{fig:circles}(a). The number of lattice points $a_m$ in $\bar{\mathcal{C}}$ is given by $|\bar{\mathcal{C}}\cap \Lambda|$ and asymptotically as $\nu_s\rightarrow 0$ (and independent of $c$)
\begin{equation}\label{eq:am}
a_m = \text{Vol}(\bar{\mathcal{C}})/\nu_s  = \frac{\omega_L}{\nu_s}\big( m^{L}-(m-1)^L).
\end{equation}

The following construction makes sure that we have $\|\lambda_1-\lambda_2\|\leq r/\sqrt{L}$. For a specific $\lambda_1\in \tilde{V}(\lambda_0)\cap\Lambda_s$ we center a sphere $\tilde{V}$ at $\lambda_1$ and use only $\lambda_2$ points from $\tilde{V}(\lambda_0)\cap\tilde{V}(\lambda_1)\cap\Lambda_s$. In Fig.~\ref{fig:circles}(b) we have shown two overlapping spheres where the first one is centered at some $\lambda_0$ and the second one is centered at some $\lambda_1\in\tilde{V}(\lambda_0)$ which is at distance $m$ from $\lambda_0$, i.e.\ $\|\lambda_0-\lambda_1\|=m/\sqrt{L}$. Let us by $\mathcal{C}$ denote the convex region obtained as the intersection of the two spheres, i.e.\ $\mathcal{C}=\tilde{V}(\lambda_0)\cap \tilde{V}(\lambda_1)$. Now let $b_m$ denote the number of lattice points in $\mathcal{C}\cap \Lambda_s$. With this we have, asymptotically as $\nu_s\rightarrow 0$, that $b_m$ is given by
\begin{equation}\label{eq:bm}
b_m = \text{Vol}(\mathcal{C})/\nu_s.
\end{equation}
\begin{figure}
\begin{center}
\psfrag{Vt}{$\tilde{V}$}
\psfrag{m}{$m$}
\psfrag{am}{$\bar{\mathcal{C}}$}
\psfrag{bm}{$\mathcal{C}$}
\psfrag{r}{$r$}
\psfrag{l0}{$\scriptstyle\lambda_0$}
\mbox{%
\subfigure[]{\includegraphics{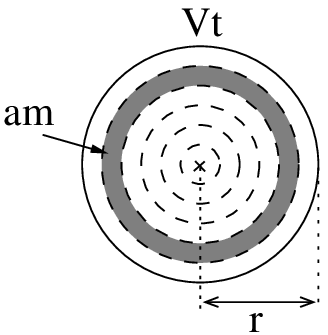}}\qquad
\subfigure[]{\includegraphics{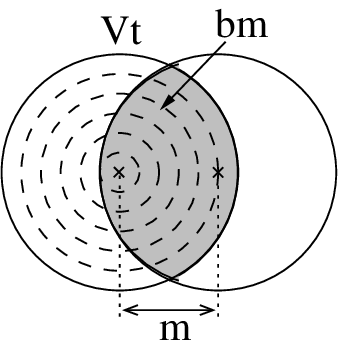}}}
\caption{The number of lattice points in the shaded region in (a) given by $a_m=\text{Vol}(\bar{\mathcal{C}})/\nu_s$ and in (b) it is given by $b_m=\text{Vol}(\mathcal{C})/\nu_s$.}
\label{fig:circles}
\end{center}
\end{figure}

It follows that the number $T$ of distinct 3-tuples which satisfy $\|\lambda_i-\lambda_j\|\leq r/\sqrt{L}$ is given by
\begin{equation}\label{eq:T}
\lim_{\nu_s\rightarrow 0} T = \sum_{m=1}^r a_m b_m.
\end{equation}

The region $\mathcal{C}$ consists of two equally sized spherical caps. We can show that the volume of an $L$-dimensional ($L$ odd) spherical cap $V_\text{cap}$ is given by (we omit the proof because of space considerations)
\begin{equation}
\begin{split}
\text{Vol}(V_\text{cap}) &= \frac{2\omega_{L-1}}{L+1} r^{(L-1)/2} (2r-m)^{(L+1)/2}\\
&\quad\times {}_2\mathcal{F}_1\left(\frac{L+1}{2},\frac{1-L}{2};\frac{L+3}{2};\frac{2r-m}{4r} \right),
\end{split}
\end{equation}
where the Hypergeometric function ${}_2\mathcal{F}_1(\cdot)$ is defined by~\cite{rainville:1960}
\begin{equation}\label{eq:hypergeom}
{}_2\mathcal{F}_1\left(a,b;c;z\right)= \sum_{k=0}^{\infty} \frac{(a)_k (b)_k}{(c)_k\, k!}z^k,
\end{equation}
where $(\cdot)_k$ is the Pochhammer symbol defined as
\begin{equation}\label{eq:pochhammer}
(a)_k=\begin{cases}
1 & k=0 \\
a(a+1)\cdots(a+k-1) & k\geq 1.
\end{cases}
\end{equation}
If either of $a$ and $b$ or both are negative, the sum in~(\ref{eq:hypergeom}) terminates. 

Inserting~(\ref{eq:am}) and~(\ref{eq:bm}) into~(\ref{eq:T}) leads to\footnote{We remark that in this asymptotical analysis we assume that all $\lambda_1$ points within a given $\bar{\mathcal{C}}$ is at exact same distance from the center of $\tilde{V}$ (i.e.\ from $\lambda_0$). The error due to this assumption is neglectable, since any constant offset from $m$ will appear inside $\mathcal{O}(\cdot)$.} (asymptotically as $\nu_s\rightarrow 0$)
\begin{equation}\label{eq:Tapprox1}
\begin{split}
T&= \sum_{m=1}^ra_m b_m \\
&=\frac{2\omega_L\omega_{L-1}}{\nu_s^2(L+1)}\sum_{m=1}^r (m^L-(m-1)^L)
r^{(L-1)/2}\\
&\quad\times (2r-m)^{(L+1)/2} {}_2\mathcal{F}_1\left(\frac{L+1}2,\frac{1-L}2;\frac{L+3}2;\frac{2r-m}{4r} \right) \\
&\overset{(a)}{=}\frac{2\omega_L\omega_{L-1}}{\nu_s^2(L+1)}r^{\frac{L-1}2}
\sum_{n=0}^{\frac{L+1}2}\binom{\frac{L+1}2}{n}(2r)^{\frac{L+1}2-n}(-1)^n \\
&\quad\times\sum_{k=0}^{\frac{L-1}2} \frac{\left(\frac{L+1}2\right)_k \left(\frac{1-L}2\right)_k}{\left(\frac{L+3}2\right)_k\, k!}\sum_{j=0}^k\binom{k}{j}\left(\frac{1}2\right)^{k-j}(-1)^j\left(\frac{1}{4r}\right)^j\\
&\quad\times\sum_{m=1}^r (m^L-(m-1)^L)m^nm^j\\
&\overset{(b)}{=} \frac{2\omega_L\omega_{L-1}}{\nu_s^2(L+1)}r^{\frac{L-1}2}
\sum_{n=0}^{\frac{L+1}2}\binom{\frac{L+1}2}{n}(2r)^{\frac{L+1}2-n}(-1)^n \\
&\quad\times\sum_{k=0}^{\frac{L-1}2} \frac{\left(\frac{L+1}2\right)_k \left(\frac{1-L}2\right)_k}{\left(\frac{L+3}2\right)_k\, k!}\sum_{j=0}^k\binom{k}{j}\left(\frac{1}2\right)^{k-j}(-1)^j\left(\frac{1}{4r}\right)^j\\
&\quad\times \left(L\sum_{m=1}^r m^{L-1+n+j} + \mathcal{O}(m^{L-2+n+j})\right).
\\
&\overset{(c)}{=} \frac{2\omega_L\omega_{L-1}}{\nu_s^2(L+1)}r^{\frac{L-1}2}
\sum_{n=0}^{\frac{L+1}2}\binom{\frac{L+1}2}{n}(2r)^{\frac{L+1}2-n}(-1)^n \\
&\quad\times\sum_{k=0}^{\frac{L-1}2} \frac{\left(\frac{L+1}2\right)_k \left(\frac{1-L}2\right)_k}{\left(\frac{L+3}2\right)_k\, k!}\sum_{j=0}^k\binom{k}{j}\left(\frac{1}2\right)^{k-j}(-1)^j\left(\frac{1}{4r}\right)^j\\
&\quad\times\left(\frac{L}{L+n+j}r^{L+n+j} + \mathcal{O}\left(r^{L-1+n+j}\right)\right),
\end{split}
\end{equation}
where $(a)$ follows by use of the binomial series expansion~\cite[p.162]{graham:1994}, i.e.\ $(x+y)^k = \sum_{n=0}^k\binom{k}{n}x^{k-n}y^{n}$, which in our case leads to 
\begin{equation}
(2r-m)^{\frac{L+1}2} = \sum_{n=0}^{\frac{L+1}2}\binom{\frac{L+1}2}{n}(2r)^{\frac{L+1}2-n}(-1)^n m^n
\end{equation}
and
\begin{equation}
\left(\frac{2r-m}{4r}\right)^k = \sum_{j=0}^k\binom{k}{j}\left(\frac{1}2\right)^{k-j}(-1)^j\left(\frac{m}{4r}\right)^j.
\end{equation}
$(b)$ is obtained  by once again applying the binomial series expansion, that is
\begin{equation}
(m-1)^L = m^L-Lm^{L-1} + \mathcal{O}(m^{L-2}),
\end{equation}
and $(c)$ follows from the fact that $\sum_{m=1}^r m^L = \frac{1}{L+1}r^{L+1}+\mathcal{O}(r^{L})$.

Next we let $r\rightarrow \infty$ so that the number of \emph{hollow} spheres inside $\tilde{V}$ goes to infinity \footnote{We would like to emphasize that this is equivalent to keeping $r$ fixed, say $r=1$, and then let the number of \emph{hollow} spheres inside $\tilde{V}$ go to infinity. To see this let $M\rightarrow\infty$ and then rewrite~(\ref{eq:am}) as
\begin{equation}\label{eq:amprime}
a_{m/M} = \text{Vol}(\bar{\mathcal{C}})/\nu_s  = \frac{\omega_L}{\nu_s}\left( \left(\frac{m}{M}\right)^{L}-\left(\frac{m-1}{M}\right)^L\right), \quad 1\leq m\leq M.
\end{equation}
A similar change applies to~(\ref{eq:bm}). Hence, the asymptotical expression for $T$ is also valid within a localized region of $\mathbb{R}^L$ which is a useful property we exploit when proving Lemma~\ref{lem:asymptsym}.}.
From~(\ref{eq:Tapprox1}) we see that, asymptotically as $\nu_s\rightarrow 0$ and $r\rightarrow \infty$, we have
\begin{equation}\label{eq:Tapprox}
T = 2\frac{\omega_L\omega_{L-1}}{\nu_s^2}\frac{L}{L+1}\beta_L r^{2L},
\end{equation}
where $\beta_L$ is constant for fixed $L$ and given by~(\ref{eq:betaL1}).

We are now in a position to find an expression for $\psi_L$. Let $\bar{\nu}$ be equal to the lower bound~(\ref{eq:vtilde}), i.e.\ $\bar{\nu}=\nu_s\sqrt{N}$ and let $\bar{r}$ be the radius of the sphere having volume $\bar{\nu}$. Then $\psi_L$ is given by the ratio of $r$ and $\bar{r}$, i.e.\ $\psi_L=r/\bar{r}$, where $r$ is the radius of $\tilde{V}$. Using this in~(\ref{eq:Tapprox}) leads to
\begin{equation}\label{eq:r}
r=\left( \frac{T\nu_s(L+1)}{2\omega_L\omega_{L-1}L\beta_L} \right)^{1/2L}.
\end{equation}
Since the radius $\bar{r}$ of an $L$-dimensional sphere of volume $\bar{\nu}$ is given by
\begin{equation}\label{eq:rbar}
\bar{r}=\left(\frac{\bar{\nu}}{\omega_L}\right)^{1/L},
\end{equation}
we can find $\psi_L$ by dividing~(\ref{eq:r}) by~(\ref{eq:rbar}), that is
\begin{equation}\label{eq:psiL1}
\psi_L=\frac{r}{\bar{r}}=\left(\frac{T\nu_s^2(L+1)}{2\omega_L\omega_{L-1}L\beta_L}\right)^{1/2L}\left(\frac{\bar{\nu}}{\omega_L}\right)^{-1/L}.
\end{equation}
Since we need to obtain $N$ 3-tuples we let $T=N$ so that with $\bar{\nu}=\sqrt{N}\nu_s$ we can rewrite~(\ref{eq:psiL1}) as
\begin{equation}\label{eq:psiL}
\psi_L=\left(\frac{\omega_L}{\omega_{L-1}}\right)^{1/2L}\left(\frac{L+1}{2L}\right)^{1/2L}\beta_L^{-1/2L}.
\end{equation}
This completes the proof. 
\end{proof}

\section{Proof of Theorem~\ref{theo:psiLinf}}\label{app:theo:psiLinf}
\begin{lemma}\label{lem:omegaratio}
For $L\rightarrow \infty$ we have
\begin{equation}
\left(\frac{\omega_{L}}{\omega_{L-1}}\right)^{1/2L}=1.
\end{equation}
\end{lemma}
\begin{proof}
The volume $\omega_L$ of an $L$-dimensional unit hypersphere is given by $\omega_L=\pi^{L/2}/(L/2)!$ so we have that
\begin{equation}
\begin{split}
\lim_{L\rightarrow \infty}&\left(\frac{\pi^{L/2}}{(L/2)!}\frac{(L/2-1/2)!}{\pi^{L/2-1/2}}\right)^{1/2L}\\
&= \lim_{L\rightarrow \infty} \pi^{1/4L} \left(\mathcal{O}(L^{-1})\right)^{1/2L}\\
&= 1.
\end{split}
\end{equation}
\rspace
\end{proof}

\begin{lemma}\label{lem:betalimit}
For $L\rightarrow \infty$ we have 
\begin{equation}
\frac{1}{\beta_L^{1/2L}} = \left(\frac{4}{3}\right)^{1/4}.
\end{equation}
\end{lemma}
\begin{proof}
The inner sum in~(\ref{eq:betaL}) may be well approximated by using that $\frac{1}{L+c}\approx \frac{1}{L}$ for $L\gg c$, which leads to
\begin{equation}\label{eq:approxfracL}
\begin{split}
&\sum_{j=0}^{k}\binom{k}{j}\left(\frac{1}{2}\right)^{k-j}(-1)^j\left(\frac{1}4\right)^j\frac{1}{L+n+j}\\
&\qquad\approx
\sum_{j=0}^{k}\binom{k}{j}\left(\frac{1}{2}\right)^{k-j}(-1)^j\left(\frac{1}4\right)^j\frac{1}{L} \\
&\qquad= \frac{1}L  \left(\frac{1}{4}\right)^k.
\end{split}
\end{equation}
We also have that 
\begin{equation}\label{eq:approxhyper}
\begin{split}
\sum_{k=0}^{\frac{L-1}2}&\frac{\left(\frac{L+1}2\right)_k \left(\frac{1-L}2\right)_k}{\left(\frac{L+3}2\right)_k\, k!} \left(\frac{1}4\right)^k\!\!=\!
{}_2\mathcal{F}_1\left(\frac{L+1}2, \frac{1-L}2;\frac{L+3}2; \frac{1}4\right) \\
&\overset{(a)}{=} (1-1/4)^{(-1+L)/2}{}_2\mathcal{F}_1\left(1, \frac{1-L}{2};\frac{L+3}2; -\frac{1}3\right) \\
&= (3/4)^{(-1+L)/2} \sum_{k=0}^{L/2-1/2} \frac{k!}{k!}\frac{(1/2-L/2)_k }{(3/2+L/2)_k} (-1/3)^k \\
&=(3/4)^{(-1+L)/2}\\
&\quad\times\sum_{k=0}^{L/2-1/2}\left(\frac{(-L/2)^k}{(L/2)^k + \mathcal{O}(L^{k-1})} + \mathcal{O}(L^{-1})\right) (-1/3)^k \\
&\approx (3/4)^{(-1+L)/2} \sum_{k=0}^{L/2-1/2} (1/3)^k,
\end{split}
\end{equation}
where $(a)$ follows from the following Hypergeometric transformation~\cite{rainville:1960}
\begin{equation}\label{eq:hypergeomtrans}
{}_2\mathcal{F}_1\left(a,b;c;z\right) = (1-z)^{-b}{}_2\mathcal{F}_1\left(c-a,b;c;\xi\right),
\end{equation}
where $\xi=\frac{z}{z-1}$. Finally, it is true that
\begin{equation}\label{eq:sumbinomn}
\sum_{n=0}^{L/2+1/2}\binom{L/2+1/2}{n} 2^{L/2+1/2-n}(-1)^n = 1.
\end{equation}
Inserting~(\ref{eq:approxfracL}),~(\ref{eq:approxhyper}) and~(\ref{eq:sumbinomn}) into~(\ref{eq:betaL1}) leads to
\begin{equation}
\beta_L \approx (3/4)^{(-1+L)/2} \frac{1}{L}\sum_{k=0}^{L/2-1/2} (1/3)^k,
\end{equation}
where since $\sum_{k=0}^{\infty} (1/3)^k = 3/2$,
we get
\begin{equation}
\begin{split}
\lim_{L\rightarrow \infty}\frac{1}{\beta_L^{1/2L}} &= \lim_{L\rightarrow \infty} (4/3)^{1/4} (4/3)^{-1/4L} L^{1/2L} (2/3)^{1/2L} \\
&= (4/3)^{1/4},
\end{split}
\end{equation}
which proves the Lemma.
\end{proof}

We are now in a position to prove the following theorem.
\begin{theo_empty}{\ref{theo:psiLinf}.}
For $K=3$ and $L\rightarrow \infty$ the dimensionless expansion factor $\psi_L$ is given by
\begin{equation}
\psi_{\infty} = \left(\frac{4}{3}\right)^{1/4}.
\end{equation}
\end{theo_empty}
\begin{proof}
The proof follows trivially by use of Lemma~\ref{lem:omegaratio} and Lemma~\ref{lem:betalimit} in~(\ref{eq:psiL}).
\end{proof}

\section{Conjecture~\ref{con:riemann2}}\label{app:riemann2}
In this appendix we justify Conjecture~\ref{con:riemann2} by proving it for the case of $K=2$ and any $L$ as well as for the case of $K=3$ and $L\rightarrow \infty$. In addition we show that it is a good approximation for the case of $K=3$ and finite $L$. 

Let $T_i = \{ \lambda_i : \lambda_i = \alpha_i(\lambda_c),\ \lambda_c \in V_\pi(0) \}$, i.e.\ the set of $N^2$ sublattice points $\lambda_i\in\Lambda_s$ associated with the $N^2$ central lattice points within $V_\pi(0)$.
Furthermore, let $T'_i \subset T_i$ be the set of unique elements of $T_i$, where $|T_i'|\approx N$. 
Finally, let $T_j(\lambda_i) = \{ \lambda_j : \lambda_j = \alpha_j(\lambda_c)\ \text{and}\ \lambda_i = \alpha_i(\lambda_c),\  \lambda_c \in V_\pi(0) \}$ and let $T'_j(\lambda_j)\subset T_j(\lambda_i)$ be the set of unique elements. That is, $T_j(\lambda_i)$ contains all the elements $\lambda_j\in \Lambda_s$ which are in the $K$-tuples that also contains a specific $\lambda_i$. We will also make use of the notation $\#_{\lambda_j}$ to indicate the number of occurrences of a specific $\lambda_j$ in $T_j(\lambda_i)$. 

For the pair $(i,j)$ we have
\begin{equation*}
\sum_{\lambda_c\in V_\pi(0)}\| \alpha_i(\lambda_c) - \alpha_j(\lambda_c)\|^2 =
\sum_{\lambda_i\in T'_i}\sum_{\lambda_j\in T_j(\lambda_i)} \|\lambda_i - \lambda_j \|^2.
\end{equation*}
Given $\lambda_i\in T'_i$, we have
\begin{equation}\label{eq:riemannassump}
\begin{split}
\sum_{\lambda_j\in T_j(\lambda_i)} \|\lambda_i - \lambda_j \|^2 \nu_s 
&=\sum_{\lambda_j\in T'_j(\lambda_i)} \#_{\lambda_j} \|\lambda_i-\lambda_j\|^2\nu_s \\
&\overset{(a)}{\approx}
\frac{N}{\tilde{N}}\sum_{\lambda_j\in T'_j(\lambda_i)}\|\lambda_i-\lambda_j\|^2\nu_s \\
&\approx \frac{N}{\tilde{N}} \int_{\tilde{V}(\lambda_i)}\|\lambda_i-x\|^2\, dx \\
&\approx \frac{N}{\tilde{N}} \tilde{\nu}^{1+2/L} G(S_L)\\
&\overset{(b)}{=} N\nu_s\tilde{\nu}^{2/L} G(S_L),
\end{split}
\end{equation}
where $(a)$ follows by assuming (see the discussion below leading to Lemma~\ref{lem:asymptsym}) that $\#_{\lambda_j}=N/\tilde{N}$ for all $\lambda_j\in T_j(\lambda_i)$ and $(b)$ follows since $\tilde{\nu}=\tilde{N}\nu_s$. Hence, with $\tilde{\nu}=\tilde{N}\nu_s=\psi N^{1/(K-1)} \nu_s$ and $\nu_s=N\nu$, we have
\begin{equation*}
\begin{split}
\sum_{\lambda_j\in T_j(\lambda_i)} \|\lambda_i - \lambda_j \|^2 \nu_s 
&\approx
N\nu_s \psi_L^{2} \nu^{2/L}N^{2/L} N^{2/L(K-1)} G(S_L)\\ 
&=
\nu_s\psi_L^{2} N^{1+2K/L(K-1)}\nu^{2/L} G(S_L),
\end{split}
\end{equation*}
which is independent of $\lambda_i$, so that
\begin{equation*}
\begin{split}
\sum_{\lambda_i\in T'_i}\sum_{\lambda_j\in T_j(\lambda_i)}\| \lambda_i - \lambda_j\|^2 
&\approx N \sum_{\lambda_j\in T_j(\lambda_i)}\| \lambda_i - \lambda_j\|^2 \\
&\approx \psi_L^{2} N^{2+2K/L(K-1)}\nu^{2/L} G(S_L).
\end{split}
\end{equation*}

In~(\ref{eq:riemannassump}) we used the approximation $\#_{\lambda_j}\approx N/\tilde{N}$ without any explanation. For the case of $K=2$ and as $N\rightarrow \infty$ we have that $T'_i = T_i$ and $N=\tilde{N}$, hence the approximation becomes exact, i.e.\ $\#_{\lambda_j}= 1$. For $K=3$ we have the following Lemma.

\begin{lemma}\label{lem:asymptsym}
For $K=3$ and asymptotically as $L\rightarrow \infty$ the following approximation becomes exact.
\begin{equation}\label{eq:g}
\sum_{\lambda_j\in T_j(\lambda_i)} \|\lambda_i-\lambda_j\|^2 \approx N\tilde{\nu}^{2/L}G(S_L).
\end{equation}
\end{lemma}

\begin{proof}
Using the same procedure as when deriving closed-form expressions for $\psi_L$ leads to the following asymptotical expression 
\begin{equation}
\sum_{\lambda_j\in T_j(\lambda_i)} \|\lambda_i-\lambda_j\|^2 = \frac{1}{L}\sum_{m=1}^r a_m b_m m^2,
\end{equation}
where we without loss of generality assumed that $\lambda_i=0$ and used the fact that we can replace $\|\lambda_j\|^2$ by $m^2/L$ for the $\lambda_j$ points which are at distance $m$ from $\lambda_i=0$. It follows that we have
\begin{equation}\label{eq:f1}
\sum_{\lambda_j\in T_j(\lambda_i)} \|\lambda_i-\lambda_j\|^2 = 2\frac{\omega_L\omega_{L-1}}{\nu_s^2}\frac{1}{L+1}\beta_L'r^{2L+2},
\end{equation}
where
\begin{equation}
\begin{split}
\beta_L'&=
\sum_{n=0}^{\frac{L+1}2}\binom{\frac{L+1}2}{n}2^{\frac{L+1}2-n}(-1)^n 
\sum_{k=0}^{\frac{L-1}2} \frac{\left(\frac{L+1}2\right)_k \left(\frac{1-L}2\right)_k}{\left(\frac{L+3}2\right)_k\, k!}\\
&\quad\times\sum_{j=0}^k\binom{k}{j}\left(\frac{1}2\right)^{k-j}(-1)^j\left(\frac{1}{4}\right)^j \frac{1}{L+n+j+2}.
\end{split}
\end{equation}

Since $\tilde{\nu}=\omega_L r^L = \psi_L^L\sqrt{N}\nu_s$ we can rewrite~(\ref{eq:f1}) as
\begin{equation}\label{eq:f}
\begin{split}
&\sum_{\lambda_j\in T_j(\lambda_i)} \|\lambda_i-\lambda_j\|^2 = 2\frac{\omega_L\omega_{L-1}}{\nu_s^2}\frac{1}{L+1}\beta_L' \tilde{\nu}^{2+2/L}\frac{1}{\omega_L^{2+2/L}} \\ 
&\quad= 2\frac{\omega_{L-1}}{\omega_L^{1+2/L}}\frac{1}{L+1}\beta_L' \tilde{\nu}^{2/L}\psi_L^{2L} N \\
&\quad\overset{(a)}{=} 2\frac{\omega_{L-1}}{\omega_L^{1+2/L}}\frac{1}{L+1}\beta_L' \tilde{\nu}^{2/L}N\left(\frac{\omega_L}{\omega_{L-1}}\right)\left(\frac{L+1}{2L}\right)\frac{1}{\beta_L} \\
&\quad=\frac{1}{\omega_L^{2/L}}\frac{1}{L}\tilde{\nu}^{2/L}N\frac{\beta_L'}{\beta_L},
\end{split}
\end{equation}
where $(a)$ follows by inserting~(\ref{eq:psiL}). Dividing~(\ref{eq:f}) by~(\ref{eq:g}) leads to
\begin{equation}
\frac{1}{\omega_L^{2/L}}\frac{1}{L}\frac{1}{G(S_L)}\frac{\beta_L'}{\beta_L}
= \frac{L+2}{L}\frac{\beta_L'}{\beta_L}.
\end{equation}
Hence, asymptotically as $L\rightarrow \infty$ we have that
\begin{equation}
\lim_{L\rightarrow \infty}\frac{L+2}{L}\frac{\beta_L'}{\beta_L} = 1,
\end{equation}
which proves the lemma.
\end{proof}

For $K>3$ it is very likely that similar equations can be found for $\psi_L$ which can then be used to verify the goodness of the approximations for any $K$. Moreover, in Appendix~\ref{app:growthriemann2} we show that the rate of growth of~(\ref{eq:riemannassump}) is unaffected if we replace $\#_{\lambda_j}$ by either $\min_{\lambda_j} \{\#_{\lambda_j}$\} or $\max_{\lambda_j} \{\#_{\lambda_j}\}$ which means that the error by using the approximation $N/\tilde{N}$ instead of the true $\#_{\lambda_j}$ is constant (i.e.\ it does not depend on $N$) for fixed $K$ and $L$. It remains to be shown whether this error term tends to zero as $L\rightarrow \infty$ for $K>3$. However, based on the discussion above we conjecture that, for any $K$, the side distortions can be expressed through the normalized second moment of a sphere as the dimension goes to infinity.

\begin{con_empty}{\ref{con:riemann2}}
For $N,L\rightarrow \infty$ and $\nu_s\rightarrow 0$, we have
for any pair $(i,j),\ i,j=0,\dots,K-1,\ i\neq j$,
\begin{equation*}
\sum_{\lambda_c\in V_\pi(0)}
\| \alpha_i(\lambda_c)-\alpha_j(\lambda_c)\|^2 = \psi_L^{2}N^{2+2K/L(K-1)}\nu^{2/L} G(S_L).
\end{equation*}
\end{con_empty}

\section{Proof of Proposition~\ref{prop:growthriemann2}}\label{app:growthriemann2}
Before proving Proposition~\ref{prop:growthriemann2} we need to lower and upper bound $\#_{\lambda_j}$ (see Appendix~\ref{app:riemann2} for an introduction to this notation).
As previously mentioned the $\lambda_j$ points which are close (in Euclidean sense) to $\lambda_i$ occur more frequently than $\lambda_j$ points farther away. To see this observe that the construction of $K$-tuples can be seen as an iterative procedure that first picks a $\lambda_0\in \Lambda_s \cap V_\pi(0)$ and then any $\lambda_1\in \Lambda_s$ is picked such that $\|\lambda_0 -\lambda_1\|\leq r/\sqrt{L}$, hence $\lambda_1\in \Lambda_s\cap \tilde{V}(\lambda_0)$.  
The set of $\lambda_{K-1}$ points that can be picked for a particular $(K-1)$-tuple e.g.\ $(\lambda_0,\dots,\lambda_{K-2})$ is then given by $\{ \lambda_{K-1} : \lambda_{K-1}\in\Lambda_s\cap \tilde{V}(\lambda_{K-2})\cap\cdots \cap \tilde{V}(\lambda_0)\}$. It is clear that $\|\lambda_i - \lambda_j\|\leq r/\sqrt{L}$ where $(\lambda_i,\lambda_j) = (\alpha_i(\lambda_c),\alpha_j(\lambda_c)), \forall \lambda_c \in \Lambda_c$ and any $i,j \in \{0,\dots,K-1\}$. 

\begin{figure}[ht]
\psfrag{l0}{$\lambda_0$}
\psfrag{l1}{$\lambda_1$}
\psfrag{l2}{$\lambda_2$}
\psfrag{C(s)}{$\mathcal{C}(s)$}
\begin{center}
\includegraphics[width=5cm]{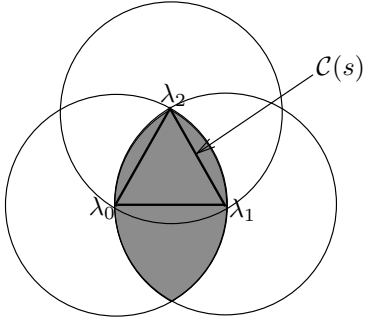}
\caption{Three spheres of equal radius are here centered at the set of points $s=\{\lambda_0,\lambda_1,\lambda_2\}$. The shaded area describes the intersection of two spheres. The equilateral triangle describes the convex hull $\mathcal{C}(s)$ of $s$.}
\label{fig:3balls}
\end{center}
\end{figure}

Let $T_\text{min}(\lambda_i,\lambda_j)$ denote the minimum number of times the pair $(\lambda_i,\lambda_j)$ is used. The minimum $T_\text{min}$ of $T_\text{min}(\lambda_i,\lambda_j)$ over all pairs $(\lambda_i,\lambda_j)$ lower bounds $N/\tilde{N}$. We will now show that $T_\text{min}$ is always bounded away from zero. To see this notice that 
the minimum overlap between two spheres of radius $r$ centered at $\lambda_0$ and $\lambda_1$, respectively, is obtained when $\lambda_0$ and $\lambda_1$ are are maximally separated, i.e.\ when $\|\lambda_0-\lambda_1\|=r/\sqrt{L}$. This is shown by the shaded area in Fig.~\ref{fig:3balls} for $L=2$. For three spheres the minimum overlap is again obtained when all pairwise distances are maximized, i.e.\ when $\|\lambda_i-\lambda_j\|=r/\sqrt{L}$ for $i,j\in \{0,1,2\}$ and $i\neq j$. 
It is clear that the volume of the intersection of three spheres is less than that of two spheres, hence the minimum number of $\lambda_2$ points is greater than the minimum number of $\lambda_3$ points. However, by construction it follows that when centering $K$ spheres at the set of points $s=\{\lambda_0,\dots,\lambda_{K-1}\} = \{\alpha_0(\lambda_c),\dots,\alpha_{K-1}(\lambda_c)\}$ each of the points in $s$ will be in the intersection $\cap_s$ of the $K$ spheres. 
Since the intersection of an arbitrary collection of convex sets leads to a convex set~\cite{rockafellar:1970},  the convex hull $\mathcal{C}(s)$ of $s$ will also be in $\cap_s$. Furthermore, for the example in Fig.~\ref{fig:3balls}, it can be seen that $\mathcal{C}(s)$ (indicated by the equilateral triangle) will not get smaller for $K\geq 3$ and this is true in general since points are never removed from $s$ as $K$ grows.
For $L=3$ the regular tetrahedron~\cite{coxeter:1973} consisting of four points with a pairwise distance of $r$ describes a regular convex polytope which lies in $\cap_s$. In general the regular $L$-simplex~\cite{coxeter:1973} lies in $\cap_s$ and the volume $V(L)$ of a regular $L$-simplex with side length $r$ is given by~\cite{buchholz:1992}
\begin{equation}
V(L)=\frac{r^L}{L!}\sqrt{\frac{L+1}{2^L}} = c_L r^L,
\end{equation}
where $c_L$ depends only on $L$. It follows that the minimum number of $K$-tuples that contains a specific $(\lambda_i,\lambda_j)$ pair is lower bounded by $V(L)^{K-2}/\nu_s^{K-2}$. Since the volume $\tilde{\nu}$ of $\tilde{V}$ is given by $\tilde{\nu}=\omega_L r^L$ we get
\begin{equation}\label{eq:lowerboundballs}
\left(\frac{V(L)}{\nu_s}\right)^{K-2} = \left(\frac{c_L}{\omega_L}\right)^{K-2}\left(\frac{\tilde{\nu}}{\nu_s}\right)^{K-2}.
\end{equation}
Also --- by construction we have that $N\leq (\tilde{\nu}/\nu_s)^{K-1}$ and that $\tilde{N}=\tilde{\nu}/\nu_s$ so an upper bound on $N/\tilde{N}$ is given by
\begin{equation}\label{eq:upperboundballs}
\frac{N}{\tilde{N}}\leq \left( \frac{\tilde{\nu}}{\nu_s}\right)^{K-2},
\end{equation}
which differs from the lower bound in~(\ref{eq:lowerboundballs}) by a multiplicative constant.

We are now in a position to prove Proposition~\ref{prop:growthriemann2}.

\begin{prop_empty}{\ref{prop:growthriemann2}}
For $N\rightarrow \infty$ and $2\leq K<\infty$ we have
\begin{equation}\label{eq:o}
\mathcal{O}\left(
\frac{\sum_{\lambda_c\in V_\pi(0)}\left\| \lambda_c - \frac{1}{K}\sum_{i=0}^{K-1}\lambda_i\right\|^2}
{\sum_{\lambda_c\in V_\pi(0)}\sum_{i=0}^{K-2}\sum_{j=i+1}^{K-1}\| \lambda_i - \lambda_j\|^2}\right) \rightarrow 0.
\end{equation}
\end{prop_empty}

\begin{proof}
The nominator describes the distance from a central lattice point to the mean vector of its associated $K$-tuple. This distance is upper bounded by the covering radius of the sublattice $\Lambda_s$. The rate of growth of the covering radius is proportional to $\nu_s^{1/L}=(N\nu)^{1/L}$, hence
\begin{equation}\label{eq:o1}
\sum_{\lambda_c\in V_\pi(0)}\left\|\lambda_c - \frac{1}K\sum_{i=0}^{K-1}\lambda_i\right\|^2= \mathcal{O}\left(N^2N^{2/L}\nu^{2/L}\right).
\end{equation}
Since the approximation $N/\tilde{N}$ used in Conjecture~\ref{con:riemann2} is sandwiched between the lower and upper bounds (i.e.\ Eqs.~(\ref{eq:lowerboundballs}) and~(\ref{eq:upperboundballs})) we can write
\begin{equation}
\begin{split}
&\sum_{\lambda_c\in V_\pi(0)}
\sum_{i=0}^{K-2}\sum_{j=i+1}^{K-1}
\|\alpha_i(\lambda_c) - \alpha_j(\lambda_c)\|^2\\
&\quad=
\sum_{i=0}^{K-2}\sum_{j=i+1}^{K-1}
\sum_{\lambda_c\in V_\pi(0)}
\|\alpha_i(\lambda_c) - \alpha_j(\lambda_c)\|^2 \\
&\quad\approx \frac{1}2K(K-1)
G(S_L)\psi_L^{2}N^2N^{2K/L(K-1)}\nu^{2/L},
\end{split}
\end{equation}
so that, since $\lambda_i=\alpha_i(\lambda_c)$,
\begin{equation}\label{eq:o2}
\sum_{\lambda_c\in V_\pi(0)}\sum_{i=0}^{K-2}\sum_{j=i+1}^{K-1}\| \lambda_i - \lambda_j\|^2
= \mathcal{O}\left(N^2N^{2K/L(K-1)}\nu^{2/L}\right).
\end{equation}
Comparing~(\ref{eq:o1}) to~(\ref{eq:o2}) we see that (\ref{eq:o}) grows as $\mathcal{O}\left(N^{-K/(K-1)}\right) \rightarrow 0$ for $N\rightarrow \infty$ and $K<\infty$.
\end{proof}


\section*{Acknowledgment}
The authors wish to thank the anonymous reviewers for their insightful and critical comments which greatly helped improving the paper.


\IEEEtriggeratref{23}
\end{document}